\documentclass{aa}
\usepackage{psfig,graphics}

\def\simgr{\mathbin{\;\raise1pt\hbox{$>$}\kern-8pt\lower3pt\hbox{$\sim$}\;}}
\def\simlr{\mathbin{\;\raise1pt\hbox{$<$}\kern-8pt\lower3pt\hbox{$\sim$}\;}}

\begin{document}



\title{Stellar and circumstellar evolution of 
       long period variable stars}

\author{M.O. Mennessier\inst{1}
        \and X. Luri\inst{2}
        }

\offprints{ M.O. Mennessier }
 
\institute{ Universit\'e de Montpellier II and CNRS, 
	    G.R.A.A.L., cc072,
            F--34095 Montpellier CEDEX 5, France
           \and 
            Departament d'Astronomia i Meteorologia,
            Universitat de Barcelona,
            Avda. Diagonal 647,
            E08028, Barcelona, Spain    
          }

\date{Received 4 May 2001; Accepted 26 September 2001}

\titlerunning{Stellar and circumstellar evolution of LPVs}
\authorrunning{M.O. Mennessier et al.}



\abstract{In a first paper, HIPPARCOS astrometric and kinematic data 
were used to
calibrate both infrared K and IRAS luminosities at the same time 
as kinematic parameters
of Long Period Variable stars (LPVs). Individual estimated absolute
magnitudes and a probabilistic assignation to galactic populations were
deduced from these calibrations for each LPV of our sample.
Here we propose a scenario of simultaneous stellar and
circumstellar evolution according to the galactic populations. The
transitory states of S and Tc stars allow us to confirm the location of
the first dredge-up at $M_{bol}=-3.5$. There is also evidence suggesting
that a previous enrichment in s-elements from a more evolved companion
may accelerate the evolution along the AGB. The possible evolution to OH
LPVs is included in this scenario, and any of these stars may have a mass 
at the limit of the capability for a C enrichment up
to $C/O > 1$. 
 A list of bright massive LPVs with peculiar envelope and luminosity properties is
 proposed as Hot Bottom Burning candidates. The He-shell flash star, R Cen, 
 is found to be exceptionally bright and could become, before leaving the AGB, 
 a C-rich LPV brighter than the usual luminosity limit of carbon stars.
}


\maketitle 


\section{Introduction} \label{sec_intro}

LPVs are particularly interesting red giants for two main reasons. On the
one hand, the brightest LPVs are luminous enough to be observed at large
distances, providing information on the host galaxy (Van
Loon et al., 1999a). On the other hand, although their precise ranges of
masses and ages remain controversial, it is clear that they are large 
enough. Thus LPVs are very good tracers of the galactic
history.
 Moreover the final evolution along the AGB, and peculiarly the carbon 
surface enrichment and the change of the envelope chemistry, is very 
complex. It depends on many factors (as convection, overshooting, 
internal chemical process, mass-loss, pulsation, etc.), the relative effects 
of which depend on the mass and metallicity, among other things.\\

In a previous paper (Mennessier et al., 2000), hereafter Paper I, HIPPARCOS
astrometric data and multi-wavelength photometric measurements of a sample
of 800 LPVs (semi-regular a and b, irregular L and Mira with O,S and C
spectral types) were analyzed using the LM algorithm (Luri, Mennessier et
al., 1996). V,K and IRAS 12 and 25 luminosities were calibrated. The LM
algorithm classified the stars according to the galactic population
(associated with the initial mass and metallicity of the stars) and to
the circumstellar envelope thickness and expansion. Several groups were
obtained in this classification:

\begin{itemize}

  \item Bright disk (BD) and disk (disk 1 or D1) LPVs
        with bright and expanding envelopes.

  \item Not so young and massive disk population (disk 2 or D2)
        divided into two subgroups: one with a thin  envelope (denoted f) and
        another one with a bright and expanding envelope (denoted b).

  \item Old disk (OD) population, showing a similar separation (b and f) according to
        envelope properties.

  \item Some LPVs were included in the extended disk (ED) population.

\end{itemize}

These groups were obtained by combining K and IRAS results.

From kinematic properties, the disk 1 population was found to be 1-4
$10^9$ yr old, disk 2 population 4-8 $10^9$ yr old and the old disk population
older than 8 $10^9$ yr, up to $10^{10}$ yr or even more. An extended disk was
assumed to be composed of very old, metal-deficient stars.

The lower limits of the main sequence initial mass, ${\cal M}_{ms}$, were
estimated to be in the range 2-1.4 ${\cal M}_{\sun}$ , 1.4-1.15 ${\cal
M}_{\sun}$, and 1.15-1 ${\cal M}_{\sun}$ for disk 1, disk 2 and old disk
populations respectively.
 Moreover, in paper I, statistical estimates were done to 
quantify how much groups 
and various variability and spectral types attract or repel 
each other.

Each star of the sample was assigned to a galactic population and its
individual K and IRAS 12 and 25 absolute magnitudes were estimated. A
table with these values is available in electronic form at CDS
\footnote {via anonymous FTP to cdsarc.u-strasbg.fr (130.79.128.5)
or via http://cdsweb.u-strasbg.fr/cgi-bin/qcat?J/A+A/}.They also 
are available in the ASTRID specialized database 
\footnote {http://astrid.graal.univ-montp2.fr}.\\

In this paper, we use the estimated individual stellar absolute magnitudes
(K) together with properties of the circumstellar envelopes (deduced from
IRAS absolute magnitudes) and the assigned galactic population to define
an evolutive scenario of simultaneous stellar and circumstellar evolution
of LPVs along the Asymptotic Giant Branch (AGB). We aimed to
link the chemical evolution from O-rich to C-rich LPVs or OH emitters
(through the intermediate states of S and/or Tc LPVs) and the stellar and
circumstellar evolution, depending on the galactic population, i.e. on the
initial mass, as discussed in sect. \ref{sec_chemev}.

In Sect. \ref{sec_gap} we examine the first stages of O-rich
LPVs and their correlation with initial mass. More precisely,  
  we propose and critically study several possible explanations 
  for the gap observed in the distribution of  O-rich LPVs, 
  separating those with and without a circumstellar shell.

  Sect. ~\ref{sec_pecstars} is dedicated to the brightest O and C-rich 
  LPVs and points out candidates for peculiarities like
  Hot Bottom Burning (HBB). Special attention is given to the case of R Cen,
  a star in a He-shell flash.

Finally, a global stellar and circumstellar evolutive scenario is proposed
in Sect.\ref{sec_conclu}, which takes into account the differences between
galactic populations and explains both chemical and variability-type
changes.

\section{Chemical evolution} \label{sec_chemev}

It is well known that K and IRAS absolute magnitudes reflect the
properties of various parts of the star. The K magnitude depends mainly
on the characteristics of the stellar surface, whereas IRAS fluxes are
linked to envelope thickness and dust composition. Using both types of
magnitudes, information about the stellar and circumstellar
properties can be obtained.

Here we mainly use the individual absolute magnitudes and
assigned galactic populations estimated in Paper I. It is convenient
to recall that {\it our sample was found to be representative of the LPV
population as far as the kinematics and the brightest luminosities are
concerned, but is under-representative for K and IRAS faint stars
} (see Paper I).

\subsection{From O-rich to C-rich LPVs} \label{sec_otoc}

\begin{figure*}

\centerline{\psfig{figure=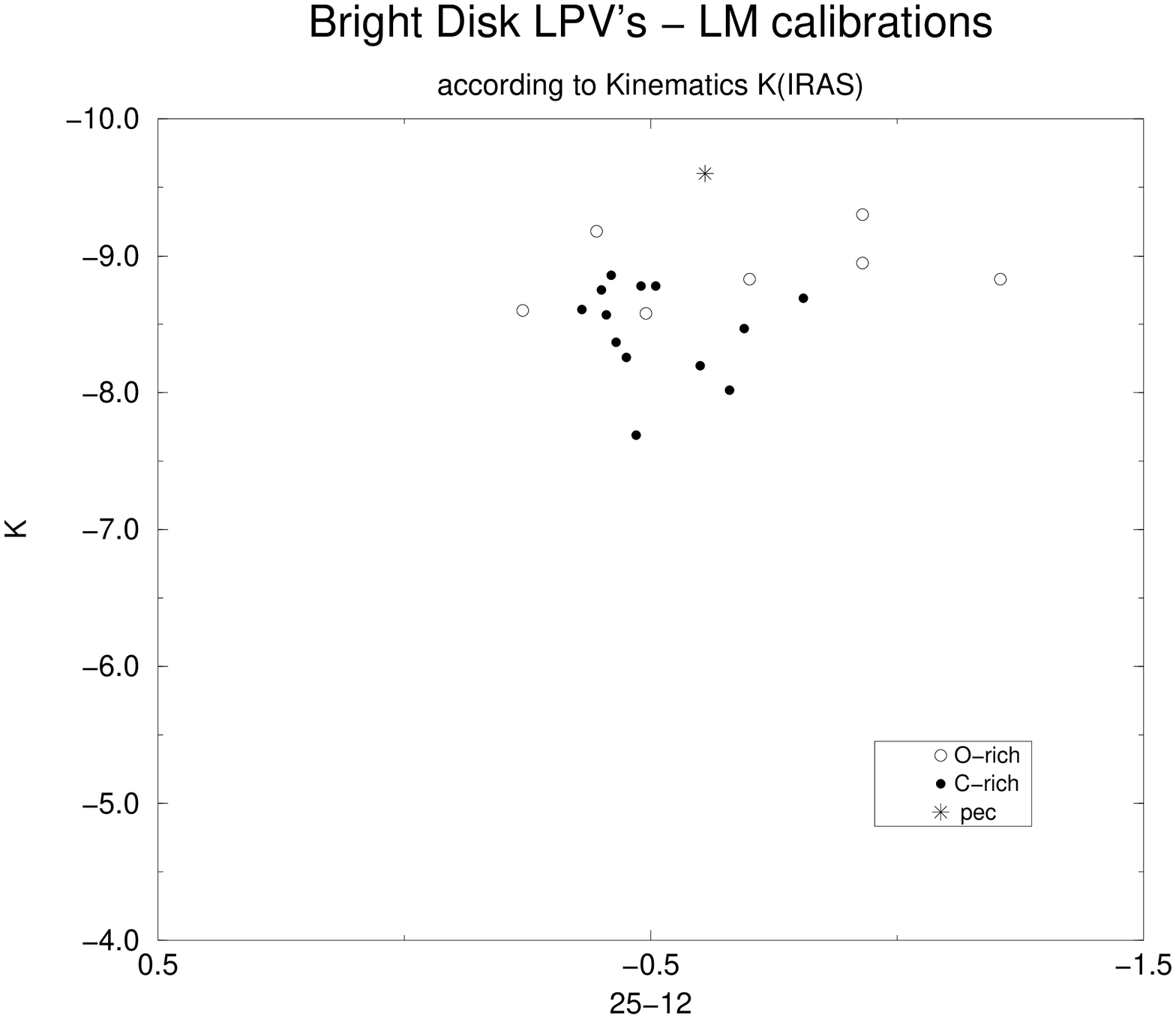,height=7cm ,angle=-90}}
\centerline{\psfig{figure=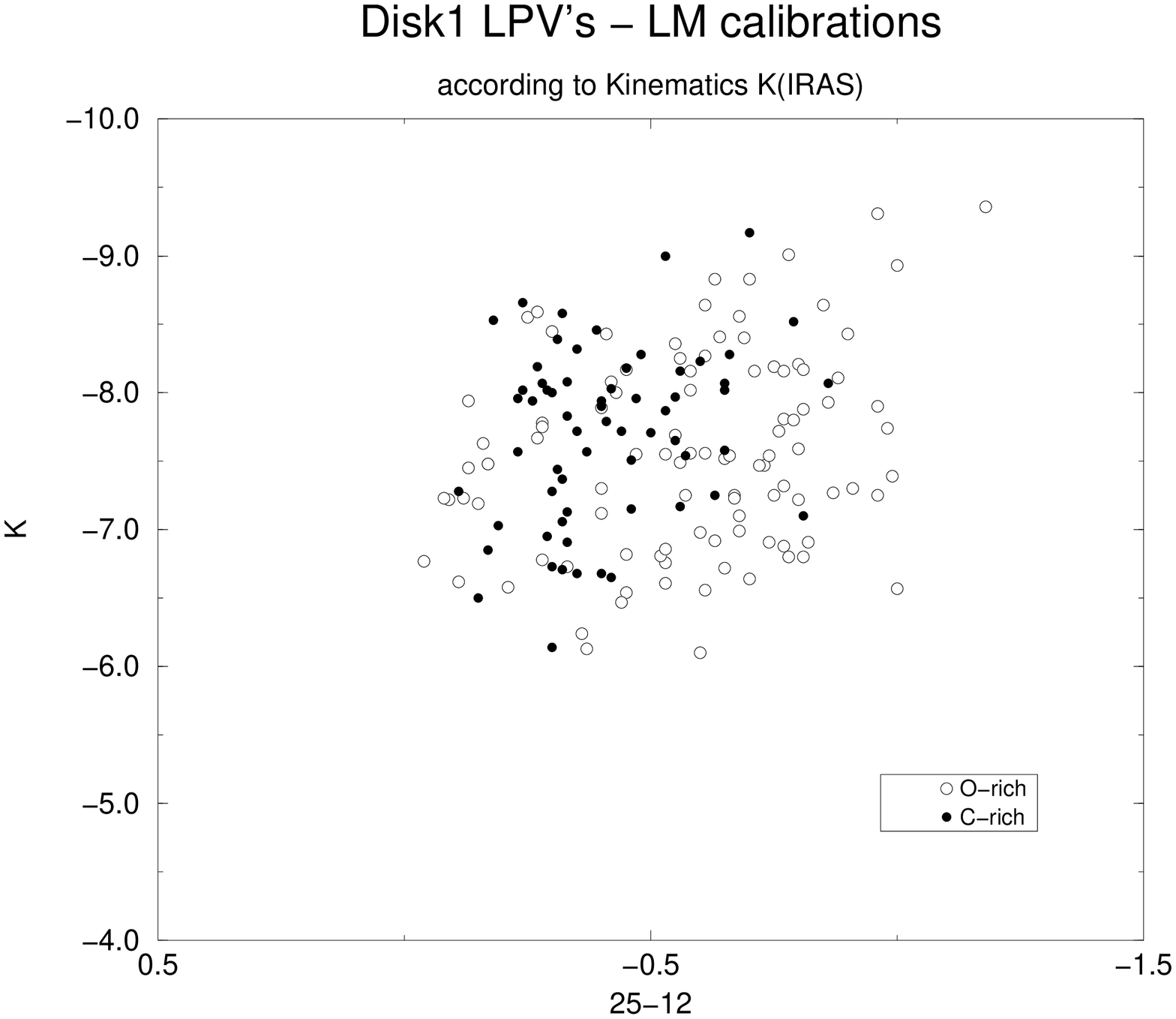 ,height=7cm ,angle=-90}
            \psfig{figure=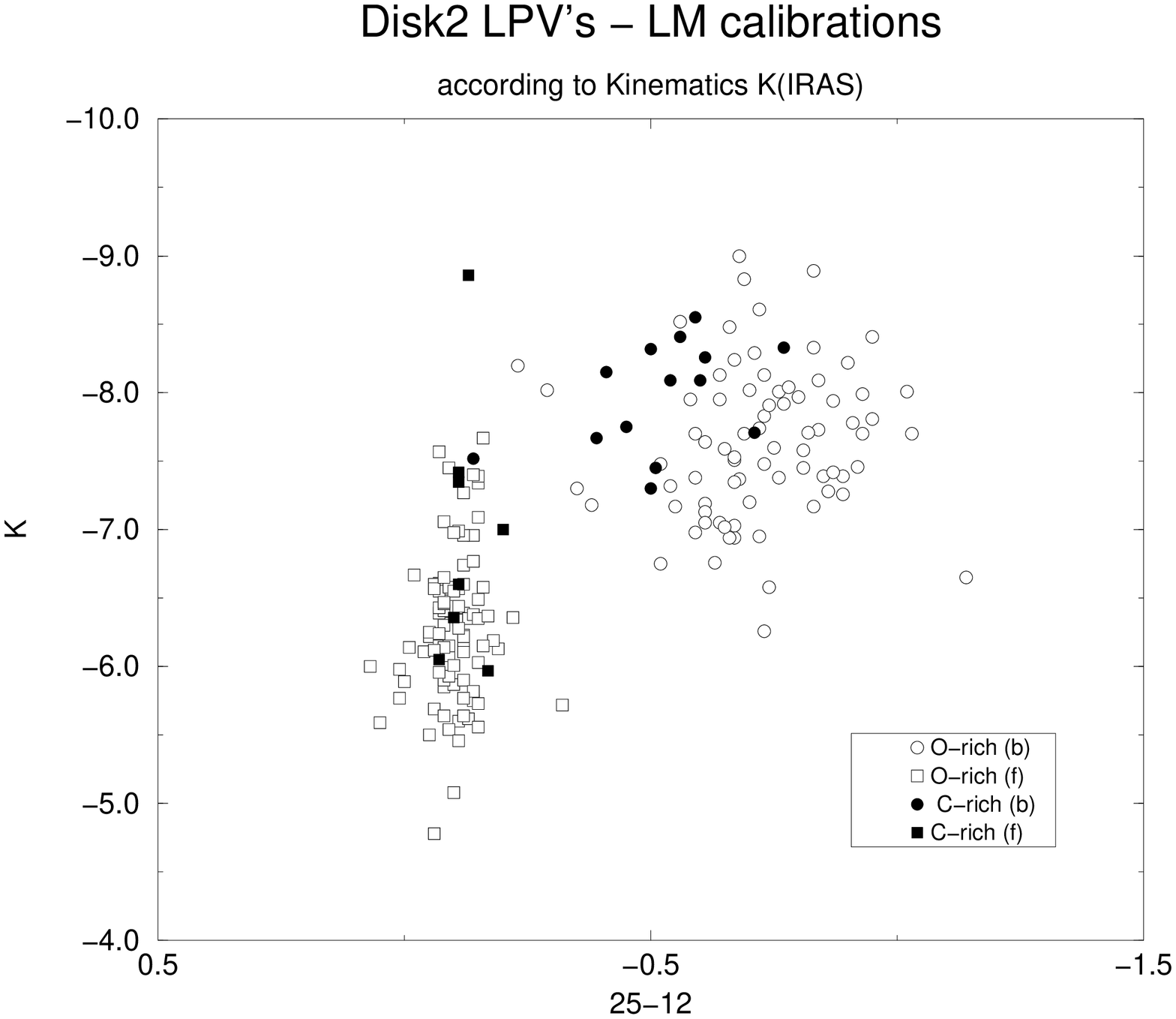,height=7cm ,angle=-90}}
\centerline{\psfig{figure=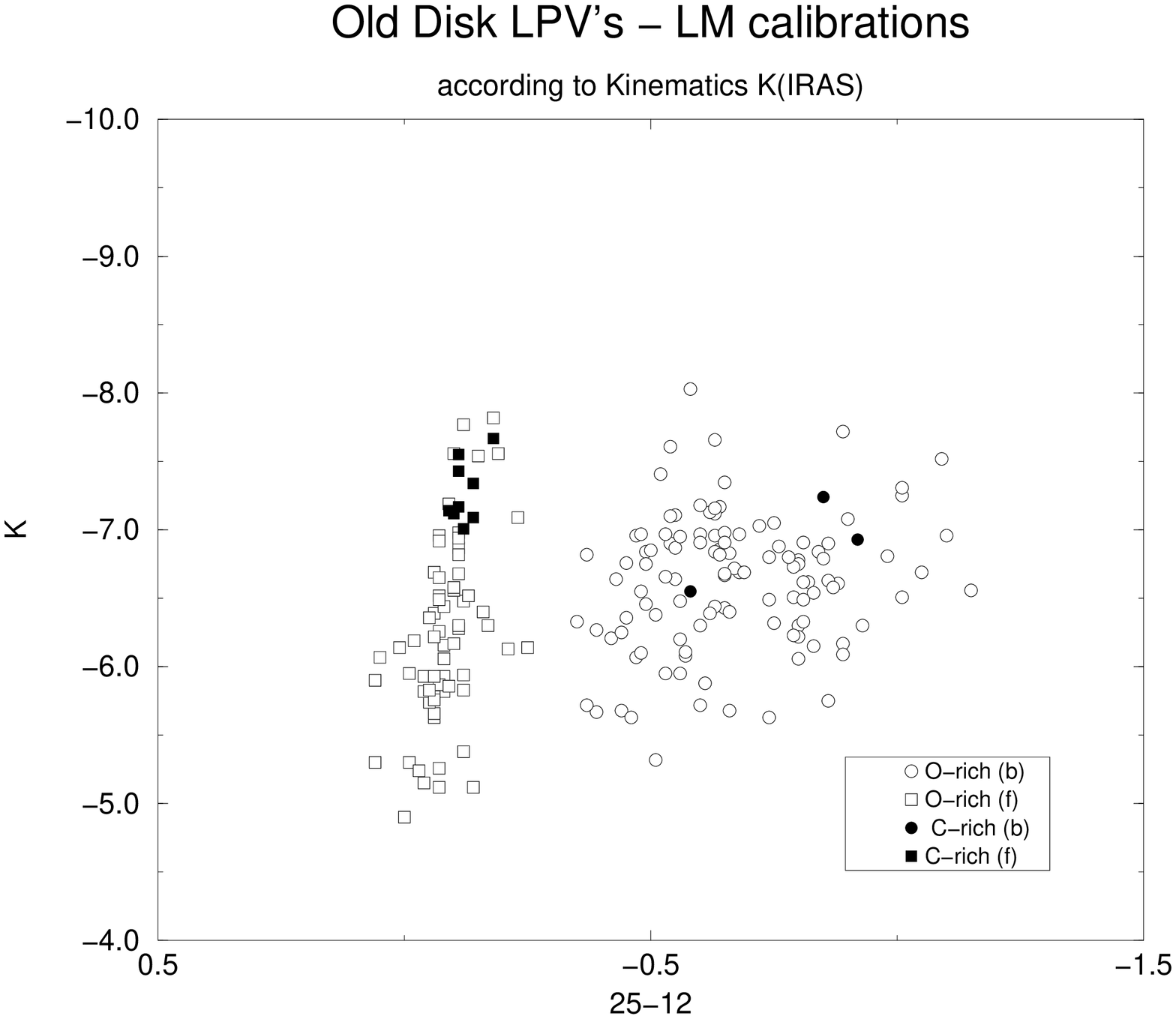 ,height=7cm ,angle=-90}
            \psfig{figure=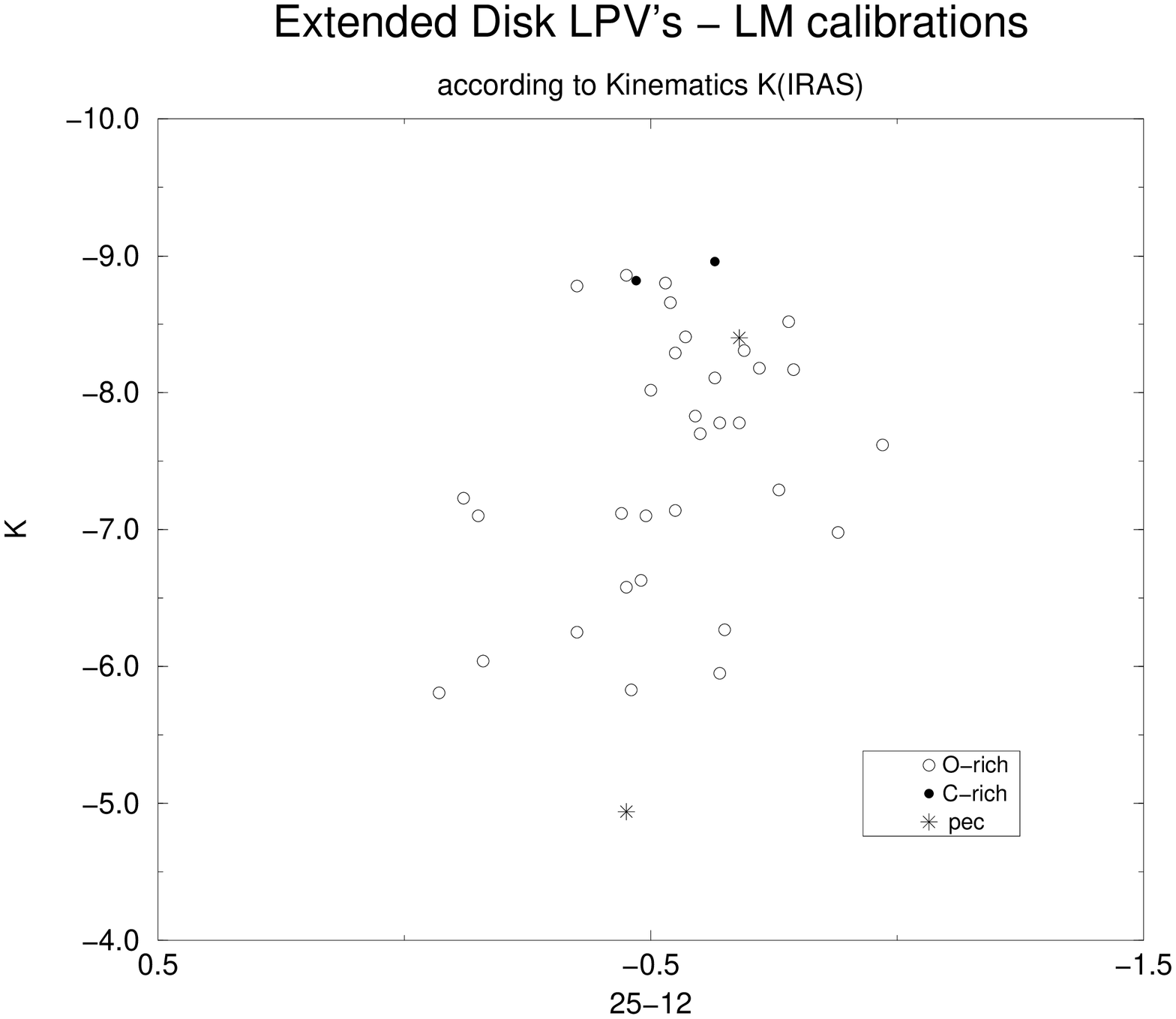,height=7cm ,angle=-90}}

\caption{Distribution of the individual estimated K luminosities and 
12-25 IRAS colors according to the assigned kinematical groups and to
spectral types.}

\label{fig_Kcolor}

\end{figure*}

Figure \ref{fig_Kcolor} shows the distribution of the individual estimated
K absolute magnitudes and 25-12 \footnote{$m_{25} = 2.07 -2.5\times \log F_{25}$ and $m_{12} = 3.63 - 2.5\times \log F_{12}$} indices deduced from
the estimated IRAS absolute magnitudes according to the assigned
kinematical groups and spectral types.   
  A bimodal distribution 
  of stars with a deficit in the number of stars
  around 25-12=-0.2 is clear, mainly for  the lower mass
  stars (disk 2 and old disk). Sect. \ref{sec_gap} will examine in detail  
  this gap for O-rich stars. In the present section we focus on the fact 
  that this area contains mainly C-rich stars belonging to disk population. 
  Moreover it corresponds to the range of the (25-12) index with the 
  lowest ratio of known variable stars among the IRAS sources with colors
  similar to LPVs colors (Paper I, fig.4).

This gap has already been reported by Habing (1989) and interpreted as the
separation between stars with and without a circumstellar envelope. The
deficit is very marked in regions I and II of  Habing's
IRAS color-color  diagram \footnote {[25]-[12] index is higher than ours 
by 1.56 mag.}, 
which correspond to non-variable and
variable (mainly O-rich) stars, and slighter in region VII, which contains
C-rich variable stars,  in agreement with our results. Thus, stars
with a thin circumstellar envelope (f) are clustered around 25-12=0 and
almost all of them are O-rich LPVs. However, the
kinematical study allowed us to assess the differences according to the
initial mass of the stars.\\

The most massive stars can evolve to C-rich LPVs after
a number of dredge-ups that enrich the external shells of the star in
Carbon. In these stars the C/O ratio becomes larger than 1 and when it
is around 1, the star is an S star.  At the same time, strong changes
take place in the circumstellar envelope, which becomes dominated by C-rich
grains. The 25-12 index increases in conformity with the loop in the IRAS
(25-12,60-25) color-color diagram predicted by Willems and de Jong (1988)
and calculated by Chan and Kwok (1988).  The areas occupied by
O-rich and C-rich disk LPVs in figure \ref{fig_Kcolor} reflect this loop.
Indeed the 25-12 index decreases from stars with a thin circumstellar
envelope (f) to O-rich LPVs with a thick envelope (b) and increases again 
for C LPVs.

\begin{figure*}

\centerline{\psfig{figure=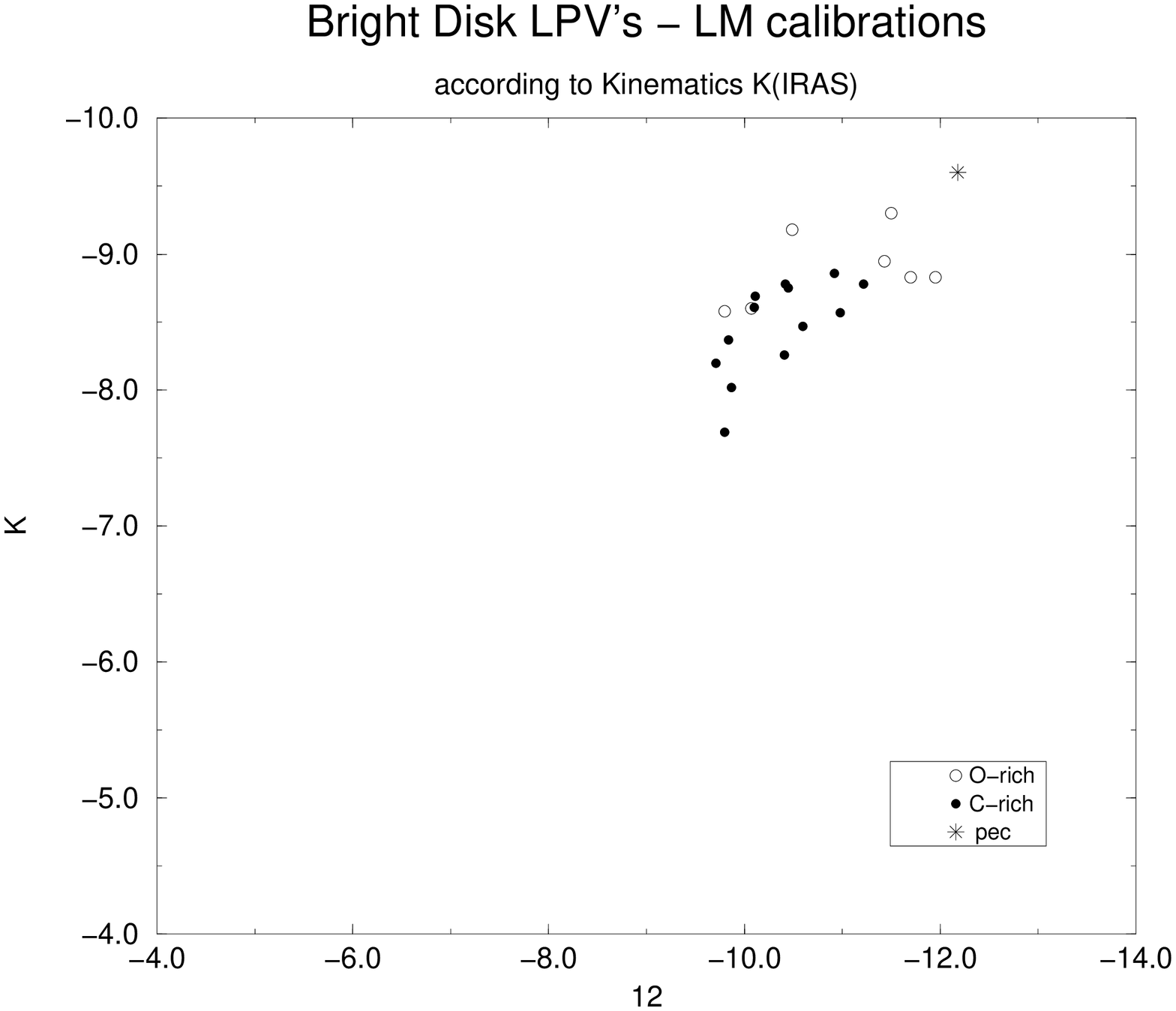 ,height=8cm ,angle=-90}}
\centerline{\psfig{figure=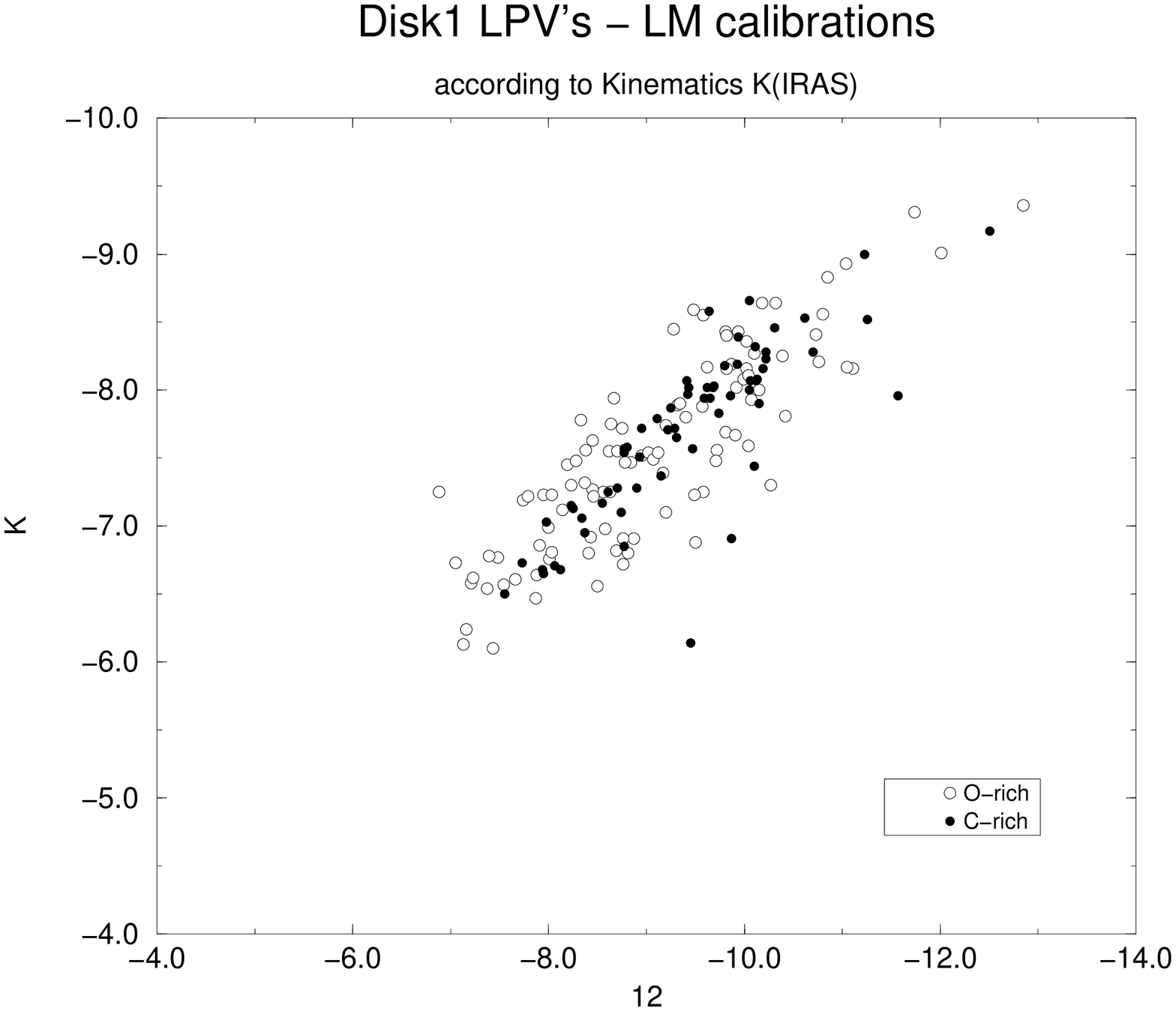 ,height=8cm ,angle=-90}
            \psfig{figure=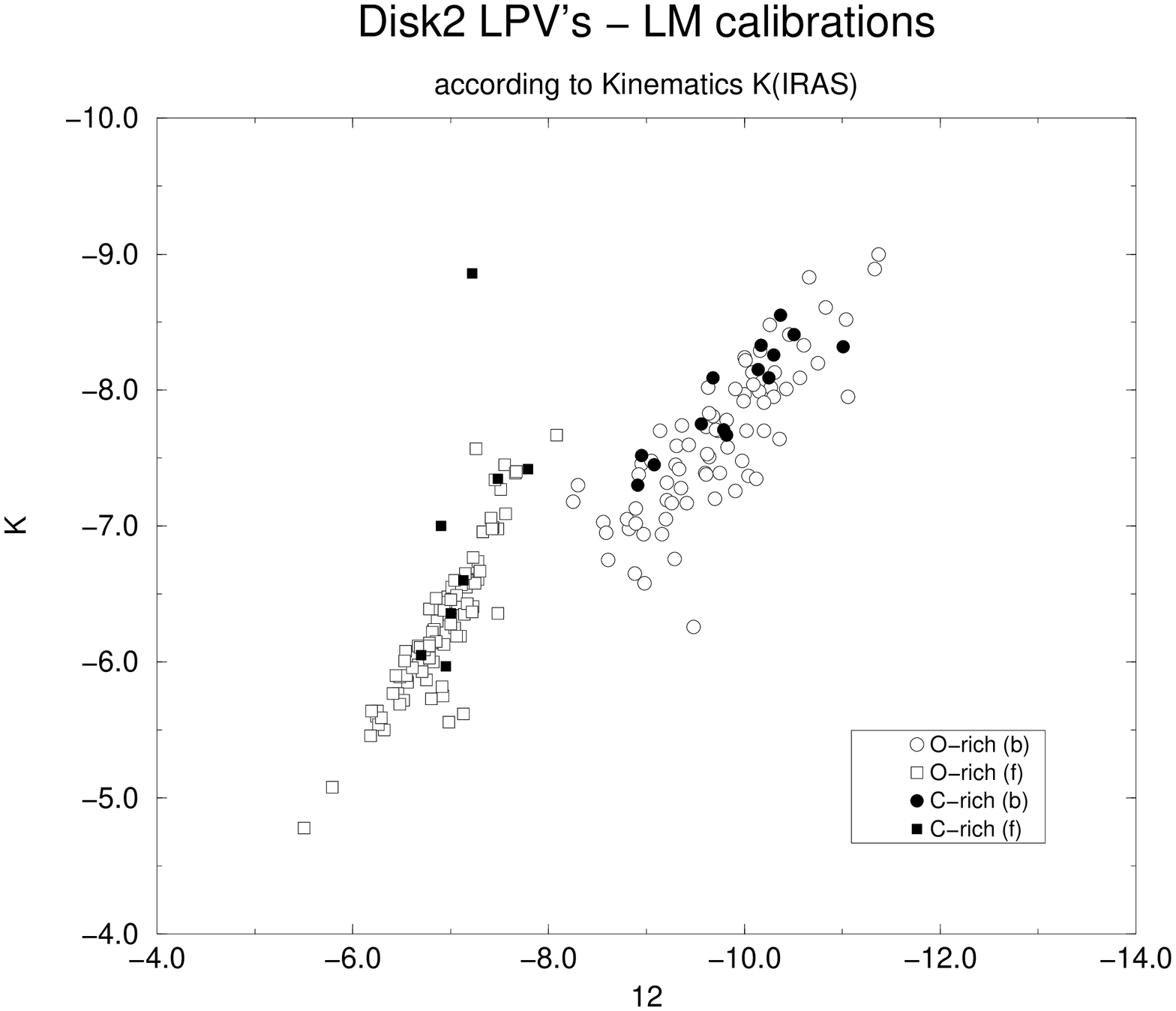 ,height=8cm ,angle=-90}}
\centerline{\psfig{figure=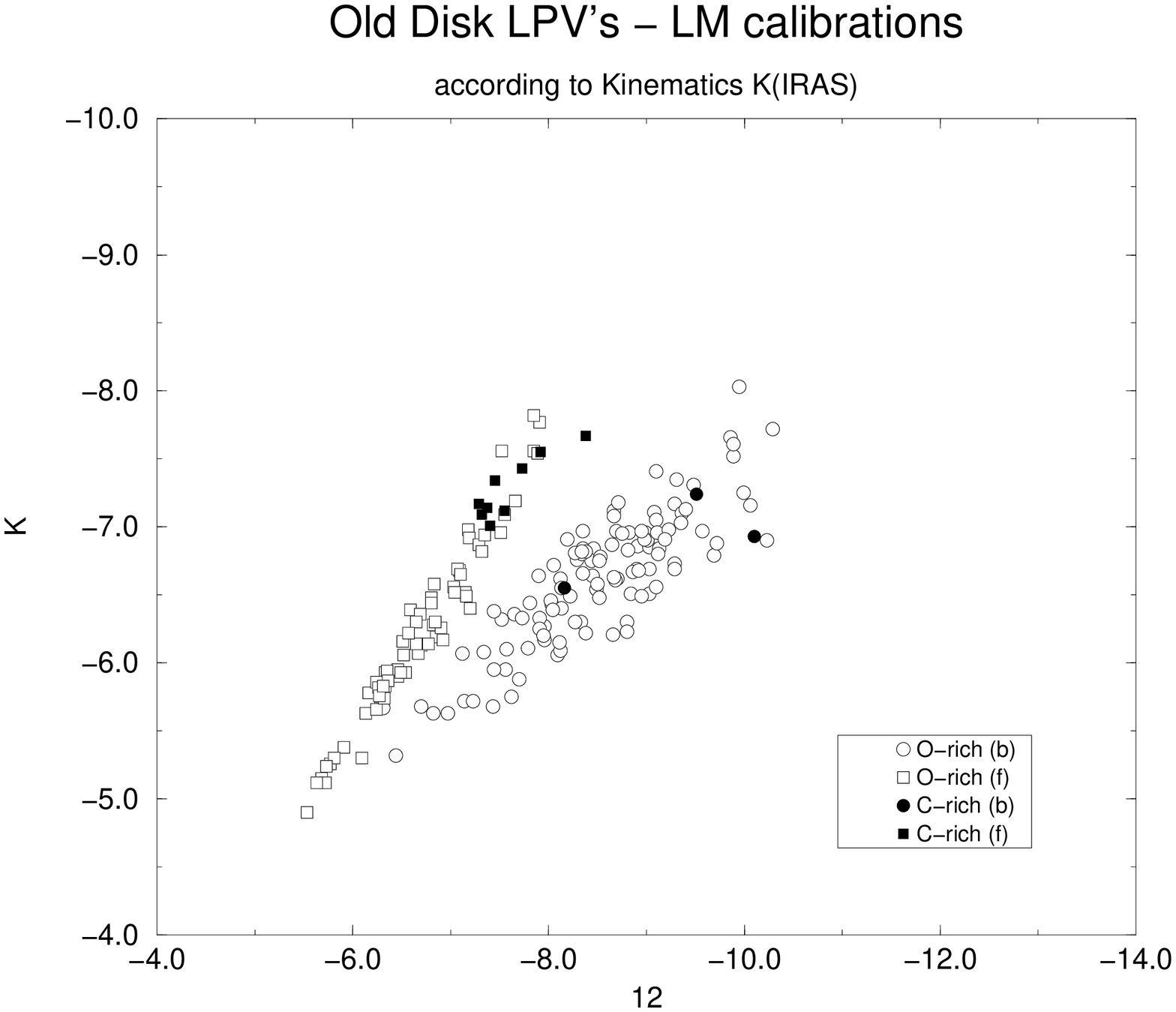 ,height=8cm ,angle=-90}
            \psfig{figure=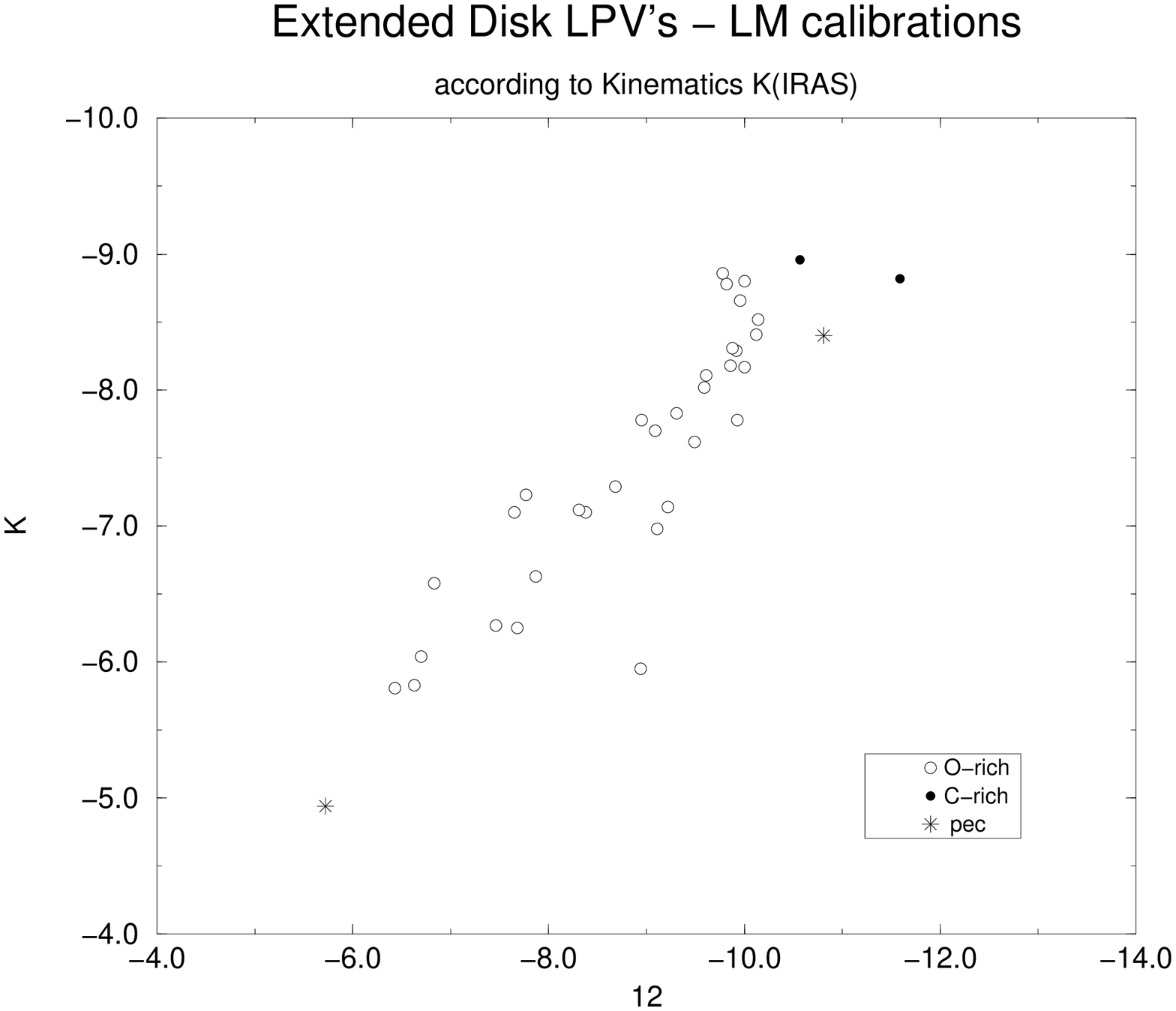,height=8cm ,angle=-90}}

\caption{Distribution of the individual estimated K and 12 luminosities 
according to the assigned kinematical groups and to
spectral types.}

\label{fig_K12}

\end{figure*}

\begin{figure*}

\centerline{\psfig{figure=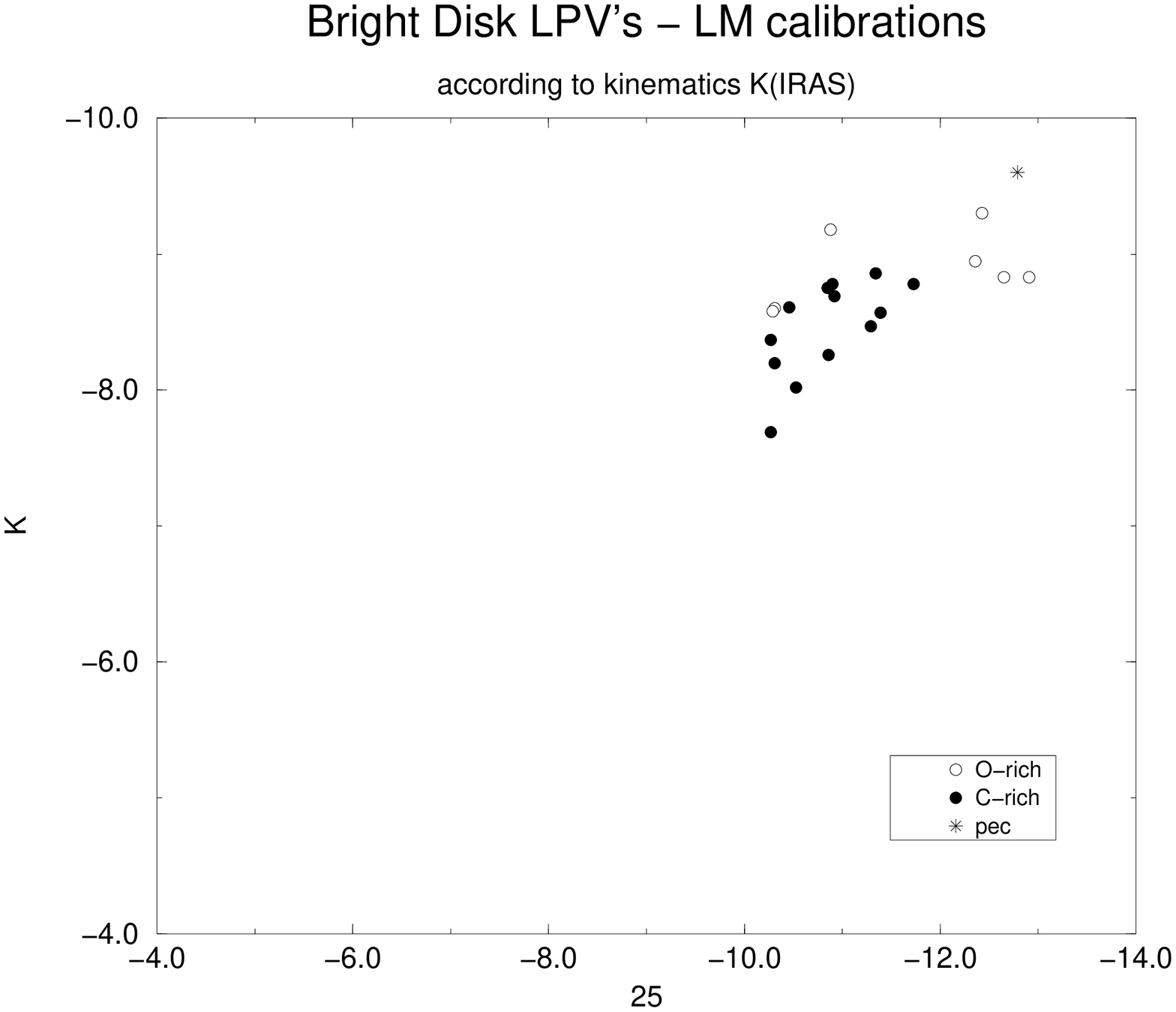 ,height=8cm ,angle=-90}}
\centerline{\psfig{figure=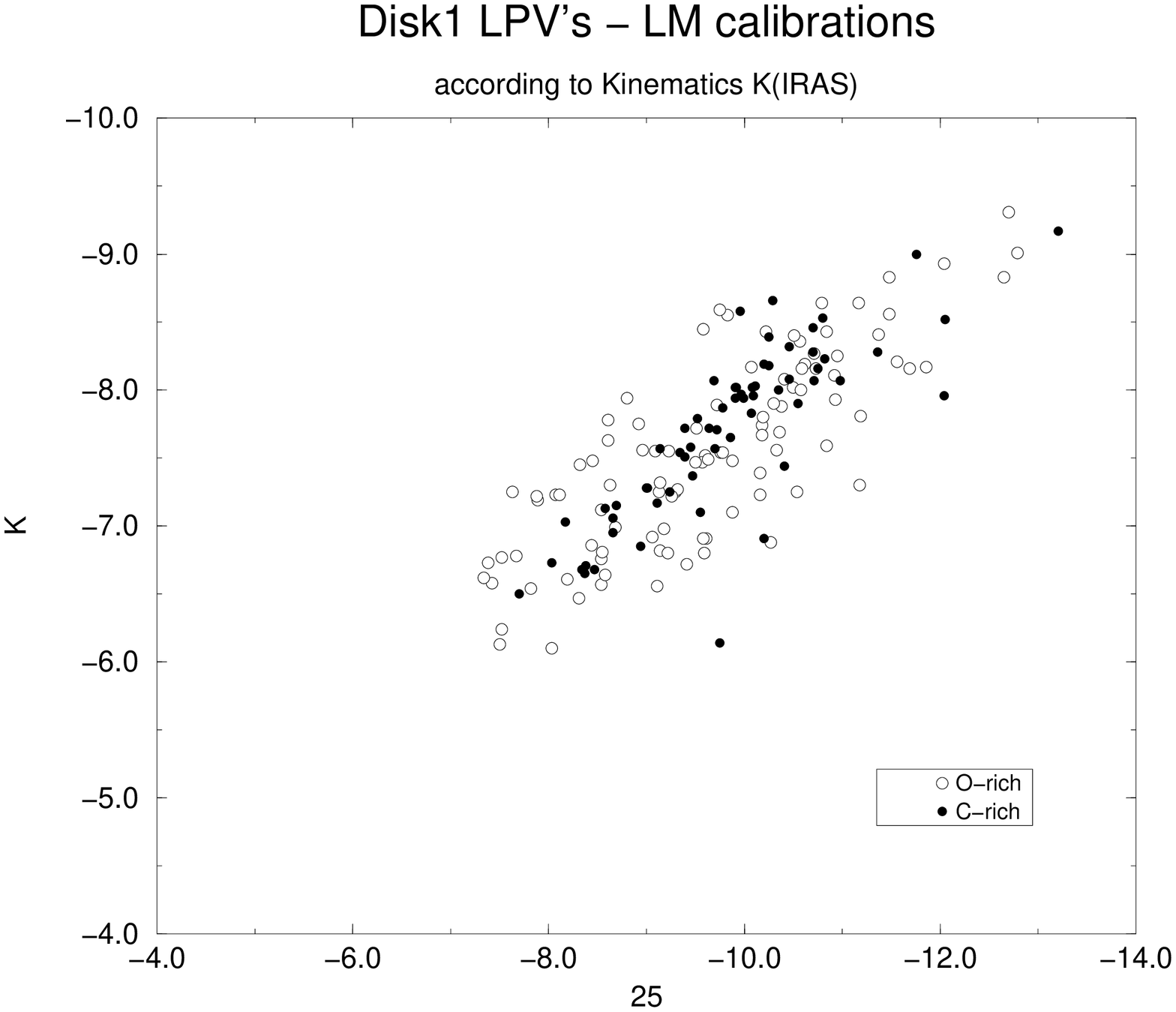 ,height=8cm ,angle=-90}
            \psfig{figure=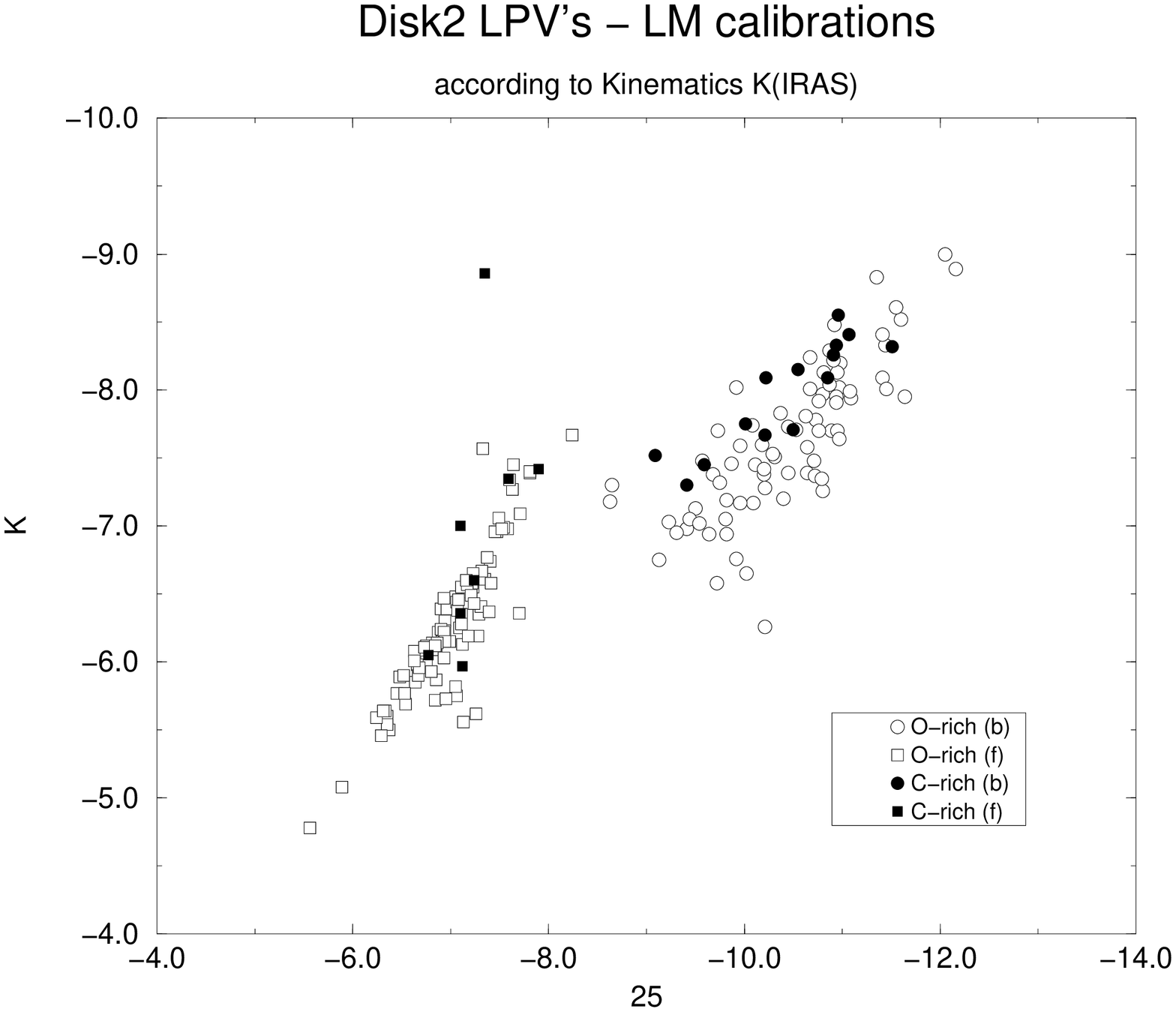 ,height=8cm ,angle=-90}}
\centerline{\psfig{figure=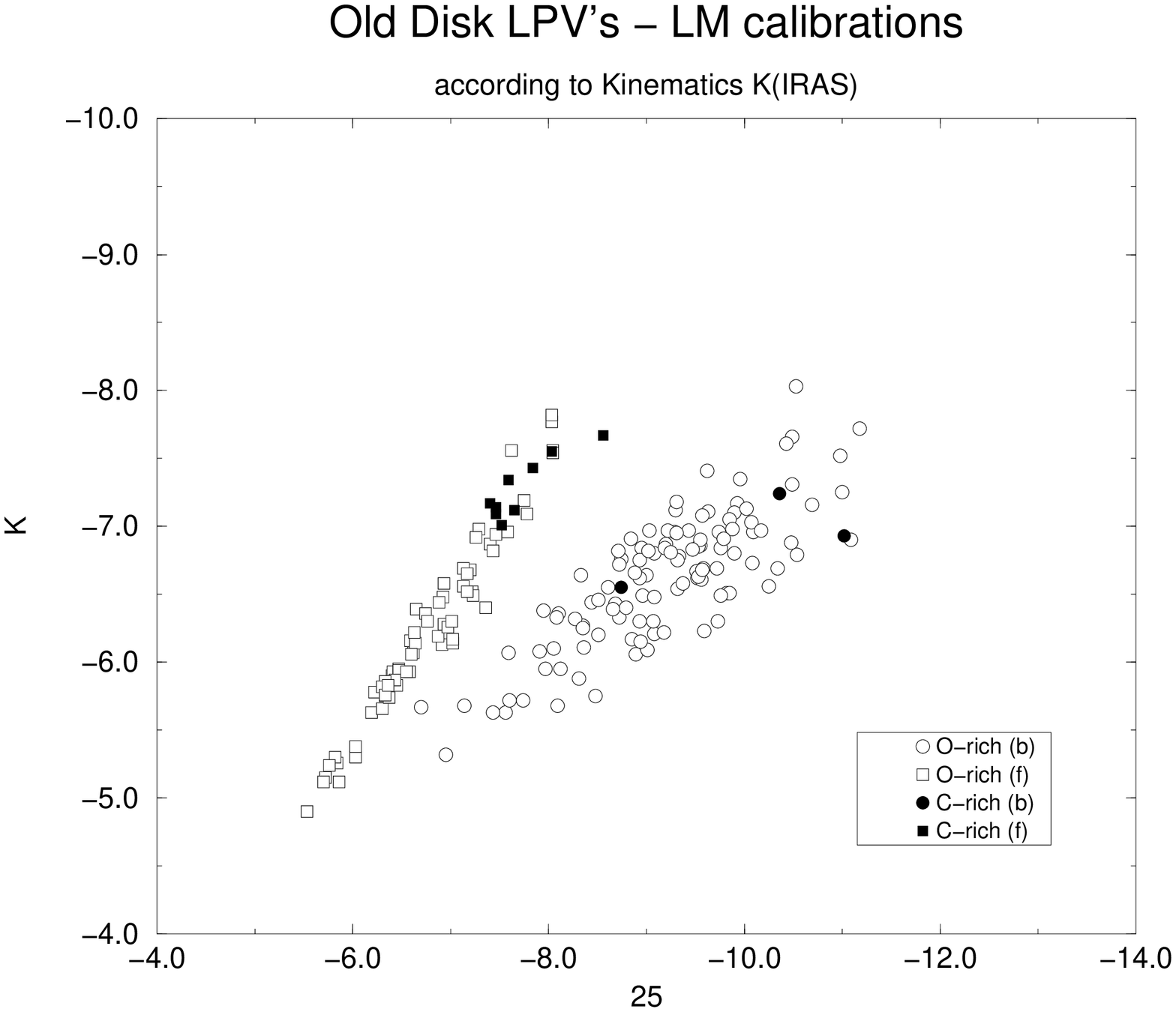 ,height=8cm ,angle=-90}
            \psfig{figure=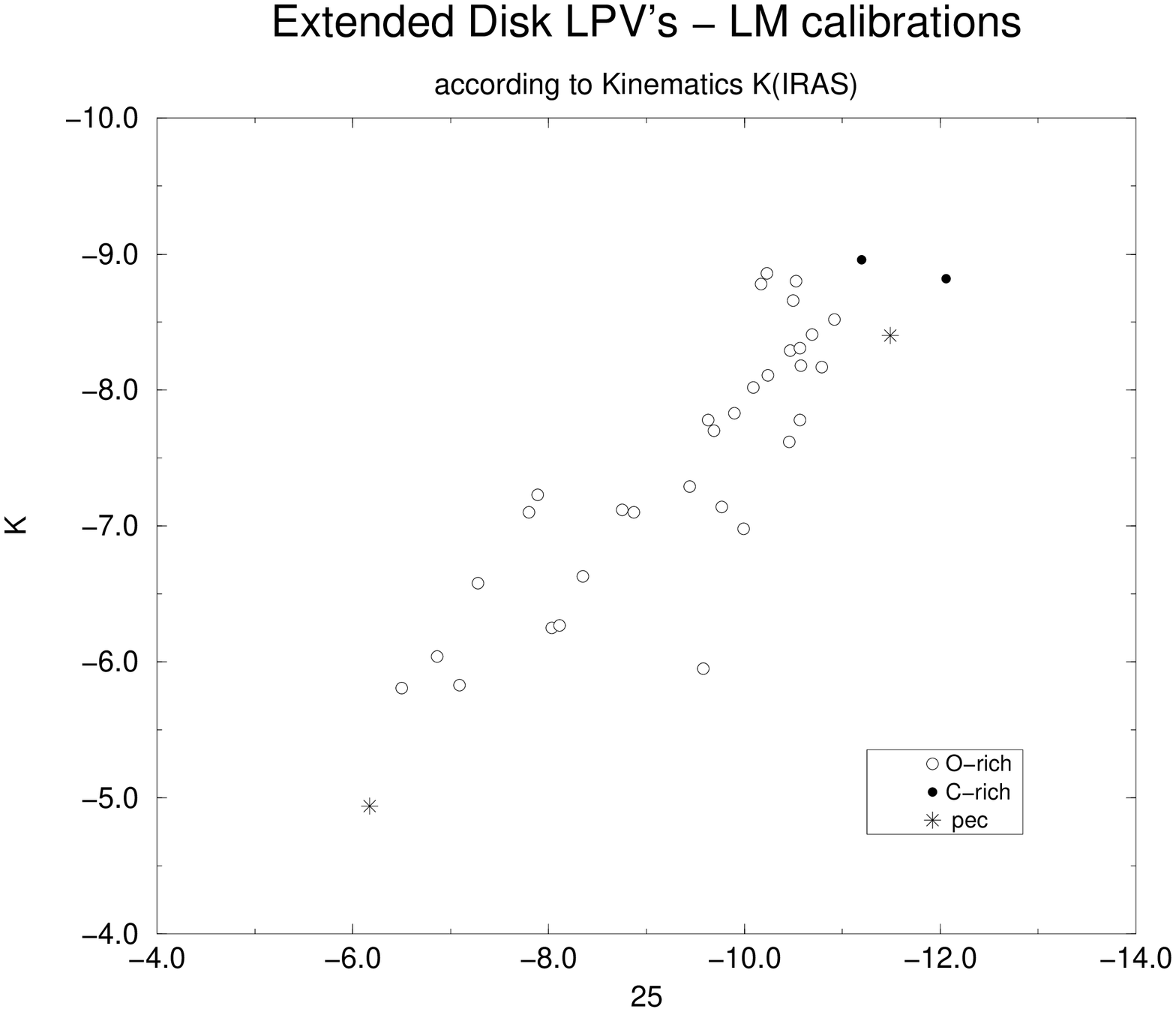,height=8cm ,angle=-90}}

\caption{Distribution of the individual estimated K and 25 luminosities 
according to the assigned kinematical groups and to
spectral types.}

\label{fig_K25}

\end{figure*}

However, our luminosity calibrations suggest that the phenomenon
does not strongly induce a difference in the K distributions of O-rich and
C-rich disk LPVs. 
Moreover, the (K,IRAS) luminosity diagrams of figures
\ref{fig_K12} and \ref{fig_K25} show that the loop is caused by both the
decreasing 12 and 25 luminosities when the star becomes C-rich, but that
this decrease is stronger in the 25 filter.\\

C-rich irregular and SRb stars mainly belong to the disk 1
population (Paper I, sect.6.4). They are mainly located at 
the upper end of the AGB. 
  Therefore, the change from a more regular variable star (Mira) 
  to a less regular one (L or SRb) 
may be associated with an increase in the non-linear behaviour of the
massive pulsating LPVs due to interactions of the pulsation phenomenon
and a very thick and dynamically unstable envelope.

\subsection{Peculiar evolutive phases} \label{sec_pecphase}

 The above global evolutive scenario along the AGB 
can be refined by examining stars which correspond to peculiar short-lived 
stages during which the star fundamentally changes (S or Tc stars), 
or to one of the most advanced stages of evolution (OH emitters).

\subsubsection{Tc stars} \label{sec_Tcstars}

According to the models, the s-elements processed in a star can be
brought to the surface by convective dredge-ups. When a sufficient
quantity of s-elements has been brought to its surface, an O-rich star
becomes an S star. S stars present a C/O ratio close to 1 and are generally
considered as a transition phase between O-rich and C-rich stars.

Some S stars are enriched in Tc, indicating that this material was brought
to the surface by recent (in the last few million years) dredge-ups, as
modeled by Mowlavi (1998) in agreement with the results obtained by van Eck et
al.(1998) who compared Tc and no-Tc S stars. Our list of
Tc-rich S LPVs is taken from van Eck's thesis (1999).\\

On another hand some O-rich LPVs (i.e. LPVs classified with 
an M spectral type) can
also be enriched in Tc. Such peculiar stars were studied by Little et al. (1987)
and they offer great potential as a possible constraint on the modelization
of dredge-ups. Table \ref{tab_Tc} shows the individual estimated 
absolute magnitudes and the assigned group of these Tc stars.

\begin{figure*}

\centerline{\psfig{figure=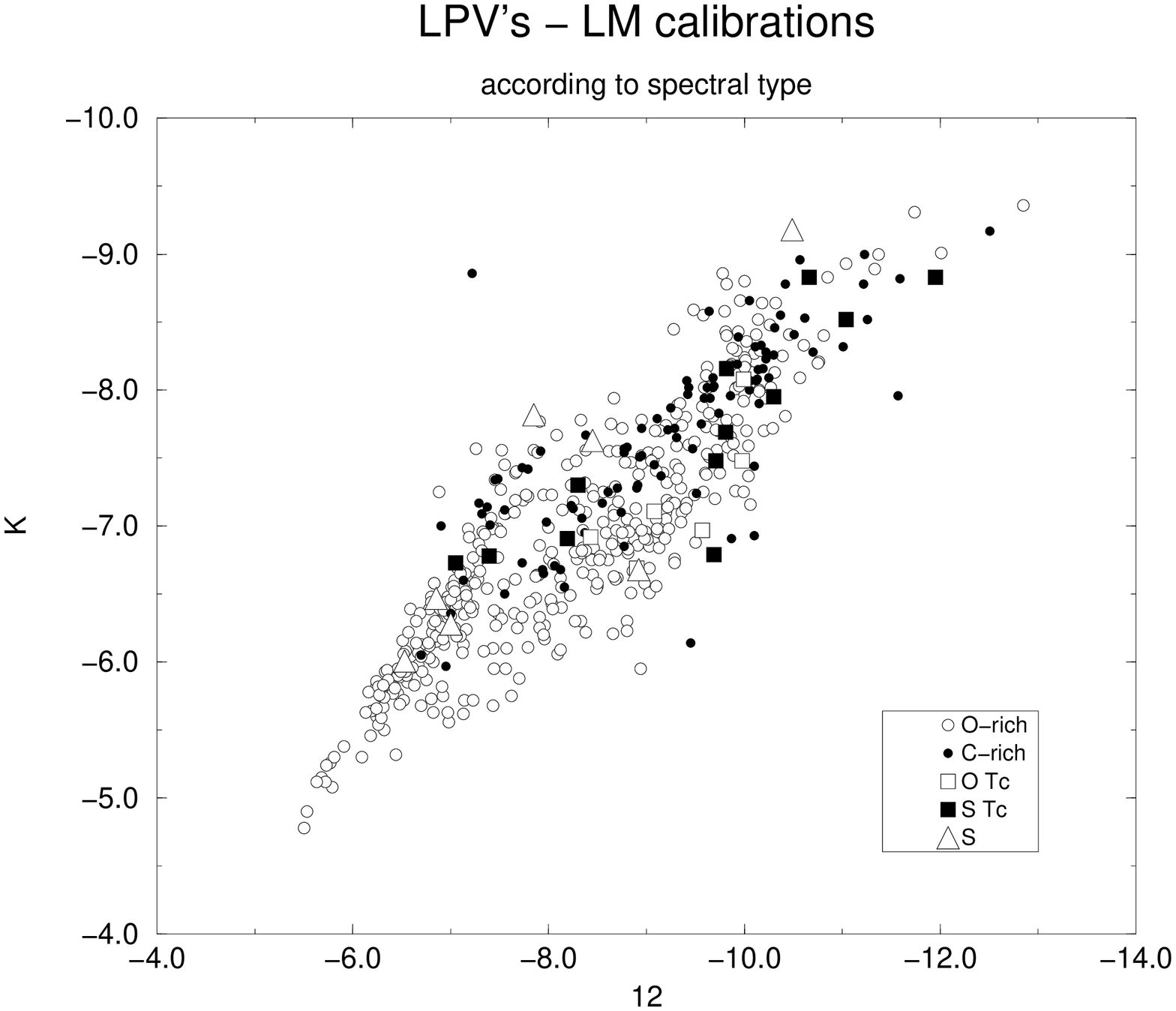,height=8cm ,angle=-90}
            \psfig{figure=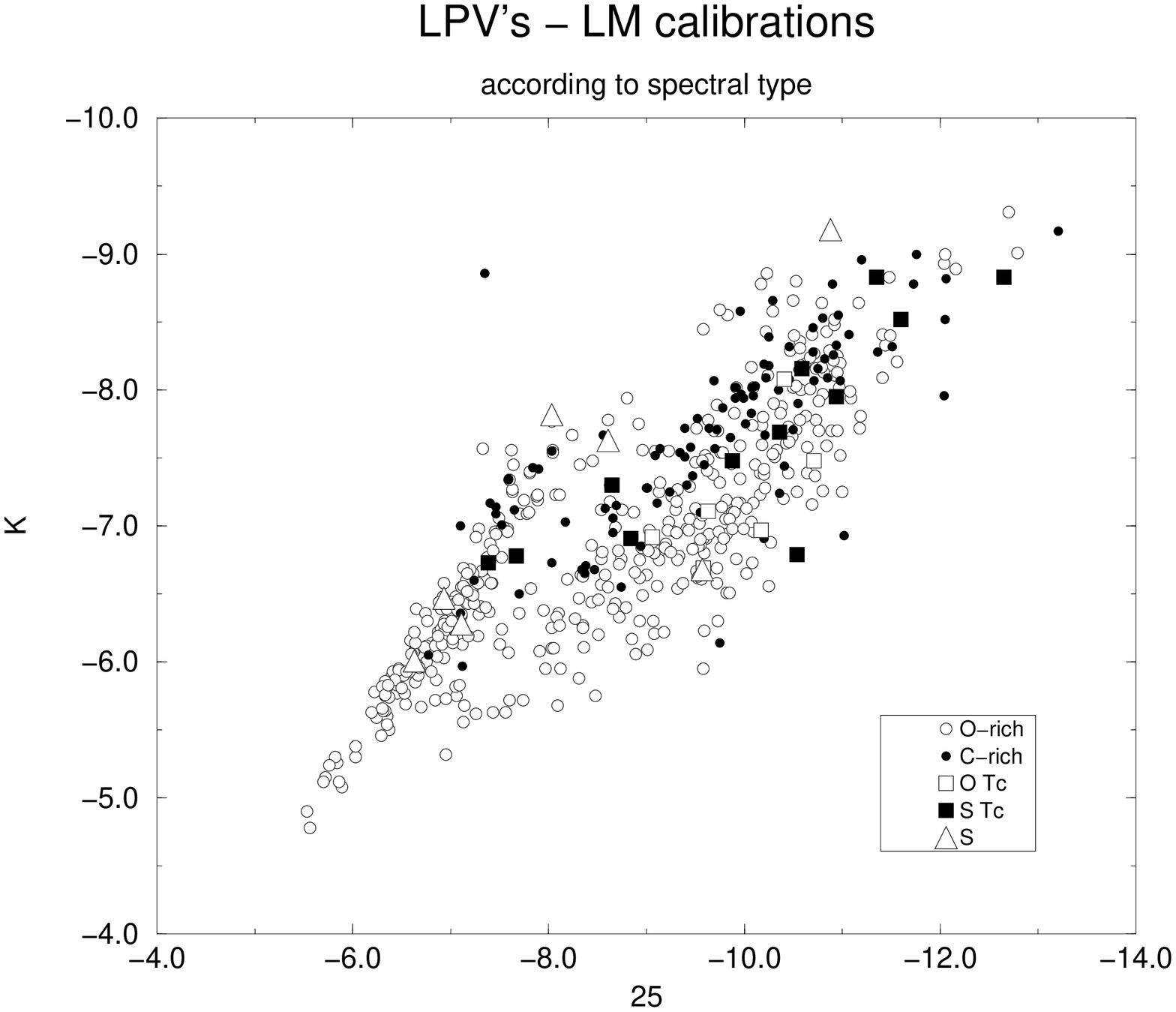,height=8cm ,angle=-90}}

\centerline{\psfig{figure=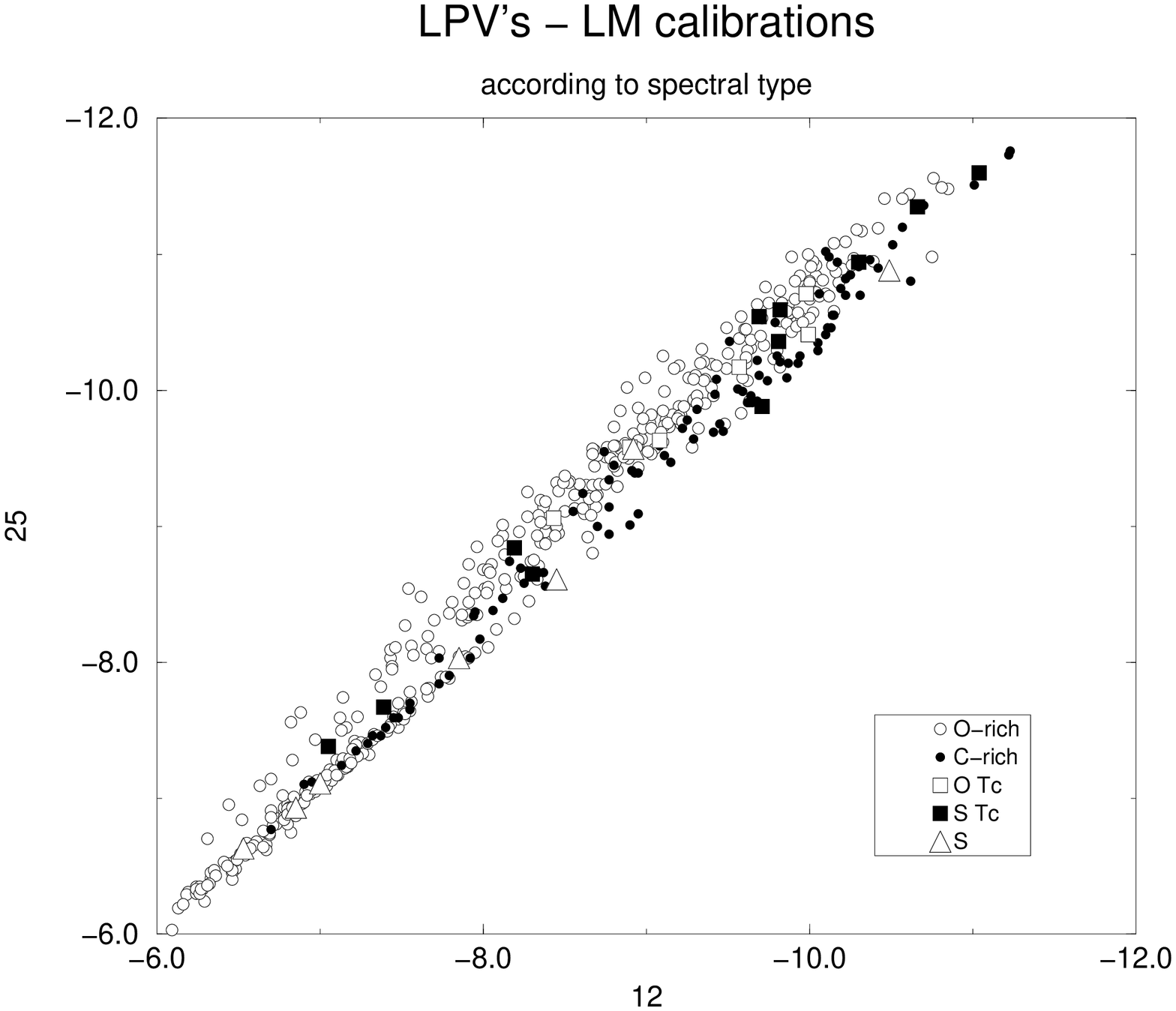,height=8cm ,angle=-90}
            \psfig{figure=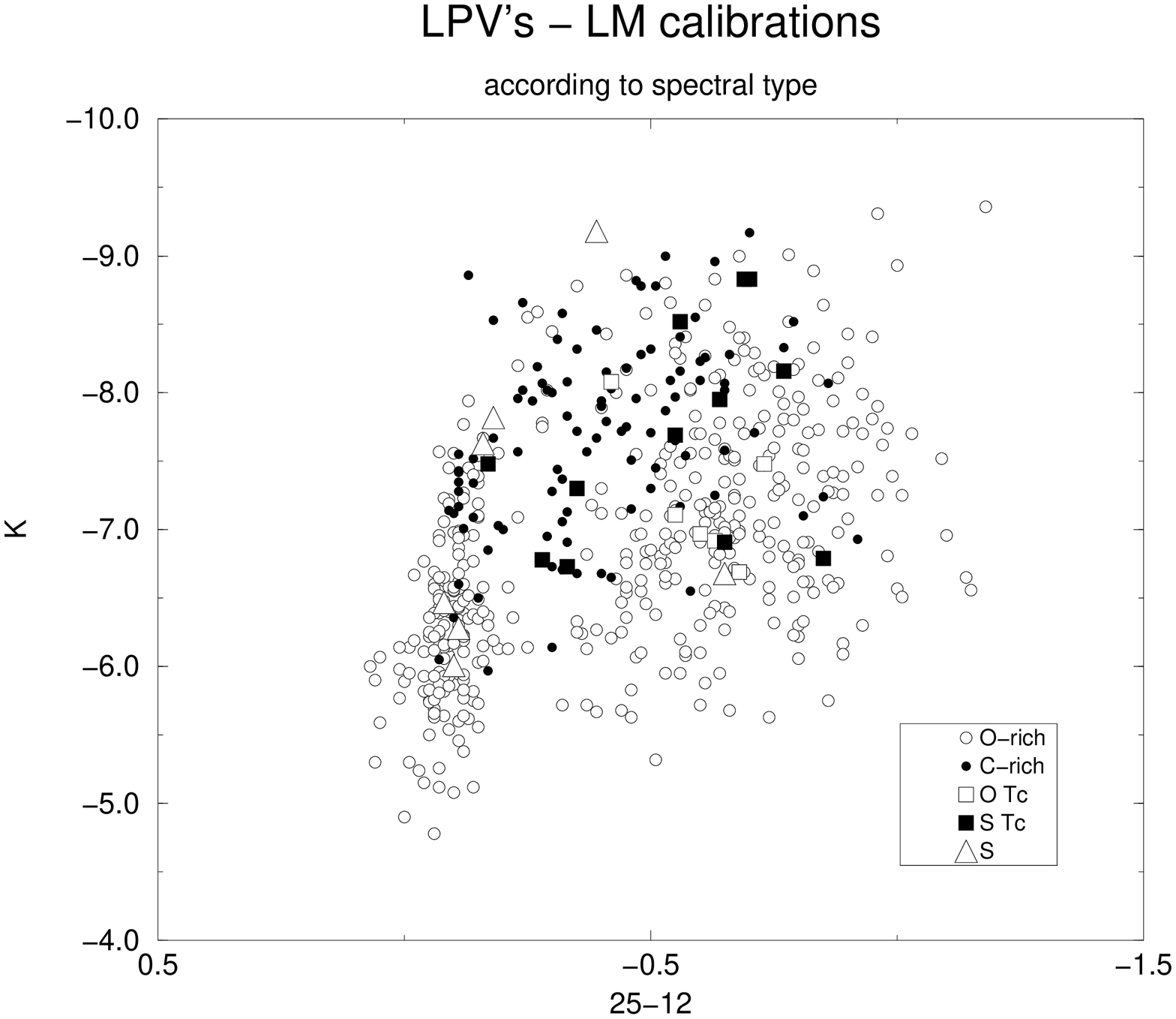,height=8cm ,angle=-90}}

\caption{Distribution of the individual estimated K, 12 and 25 luminosities and 
IRAS colors of Tc and S stars compared with O and C-rich LPVs.
}

\label{fig_Tc}

\end{figure*}

\begin{table*}

\caption{Individual K,12 and 25 luminosities, with assigned crossing K(IRAS) 
group of Tc O-rich and S spectral type LPVs. 
Variability (M=Mira,SR=semi regular,L=irregular) types
and possible specificity (Tc=Technetium star,  BD=bright galactic disk star)
are given.}
\label{tab_Tc}
\begin{center}
\begin{tabular}{|l|rr|r|r|rrr|r|}
   \hline
HIP id & \multicolumn{2}{c}{name} & types & group & K & 12 & 25 & pec \\
   \hline

     8 & Z     & Peg &    MO &  ODb &   -6.69 &   -8.90 &   -9.58 & Tc \\ 
  1236 & S     & Scl &    MO &  ODb &   -7.11 &   -9.08 &   -9.63 & Tc \\ 
 77615 & R     & Ser &    MO &  ODb &   -6.97 &   -9.57 &  -10.17 & Tc \\ 
 90493 & RV    & Sgr &    MO &    D1 &   -6.92 &   -8.43 &   -9.06 & Tc \\ 
104451 & T     & Cep &    MO &     D1 &   -8.08 &   -9.99 &  -10.60 & Tc \\ 
110736 & S     & Gru &    MO &   D2b &   -7.48 &   -9.98 &  -10.71 & Tc \\ 
      &        &     &       &       &         &         &         &   \\ 
  1901 & R     & And &    MS &  ODb &   -6.79 &   -9.69 &  -10.54 & Tc \\ 
 10687 & W     & And &    MS &     D1 &   -8.83 &  -11.95 &  -12.65 & Tc,BD \\
 34356 & R     & Gem &    MS &     D1 &   -7.69 &   -9.81 &  -10.36 & Tc \\ 
 35045 & AA    & Cam &    LS &  ODb &   -6.91 &   -8.19 &   -8.84 & Tc \\ 
 38502 & NQ    & Pup &    LS &   D2f &   -6.73 &   -6.97 &   -7.05 & Tc \\ 
 65835 & R     & Hya &    MS &   D2b &   -8.52 &  -11.04 &  -11.60 & Tc \\ 
 87850 & OP    & Her &   SRS &   D2b &   -7.30 &   -8.30 &   -8.65 & Tc \\ 
 94706 & T     & Sgr &    MS &   D2b &   -7.95 &  -10.30 &  -10.94 & Tc \\ 
 97629 & khi   & Cyg &    MS &     D1 &   -7.48 &   -9.71 &   -9.88 & Tc \\ 
 98856 & AA    & Cyg &   SRS &   D2b &   -8.83 &  -10.66 &  -11.35 & Tc \\ 
110478 & pi.1  & Gru &   SRS &     D1 &   -8.16 &   -9.82 &  -10.59 & Tc \\ 
113131 & HR    & Peg &   SRS &     D1 &   -6.78 &   -7.39 &   -7.67 & Tc \\ 
      &        &     &       &       &         &         &         &   \\ 
 17296 & BD    & Cam &    LS &   D2f &   -6.01 &   -6.53 &   -6.63 & \\ 
 33824 & R     & Lyn &    MS &  ODf &   -7.82 &   -7.85 &   -8.03 & \\ 
 40977 & V     & Cnc &    MS &   D2f &   -6.47 &   -6.85 &   -6.93 & \\ 
101270 & AD    & Cyg &    LS &     D1 &   -9.18 &  -10.49 &  -10.88 & BD \\ 
110146 & X     & Aqr &    MS &  ODb &   -6.68 &   -8.92 &   -9.57 & \\
112784 & SX    & Peg &    MS &   D2f &   -6.28 &   -7.00 &   -7.11 & \\ 
114347 & GZ    & Peg &  SRSa &     D1 &   -7.63 &   -8.45 &   -8.61 & \\ 
  \hline
  \end{tabular}
  \end{center}
\end{table*}

We would like to highlight the 
correlation between location in the plane (12,25) of Tc LPVs and the limit
between O-rich and C-rich regions, regardless of their S or M spectral
type (see figure \ref{fig_Tc}).  The only exception is R And, which will be
discussed later in this paper.\\

However no differences are found between Tc O-rich and Tc S LPVs.
Tc O-rich LPVs have the same 12 and 25 luminosities as
O-rich stars. They are probably LPVs enriched in Tc by a recent dredge-up,
but not efficient enough either to make the C/O ratio close to 1 or to
drastically alter the circumstellar envelope. 
Tc S LPVs are mainly assigned to disk population (10/12), which is not
valid for Tc O-rich LPVs. This suggests that the dredge-up is
more efficient in changing the C/O surface ratio up to 1 for more massive
stars.  No definitive conclusions can be reached  owing to
the scarcity of Tc LPVs in the sample.\\

  It is also important to note  
that all these stars are more luminous than $\simeq$ -6.5 mag in K, in 
agreement with  Van Eck et al. (1998), the bolometric correction for this
type of stars being around 3 mag. This confirms the predicted location of
the first thermal pulse (Mowlavi, 1998) and the quite early operative
third dredge-up on the TP-AGB (Van Eck, 1999). \\

 \noindent Finally, some individual Tc stars in our sample have specific 
properties that require a specific discussion:

\begin{itemize}

  \item NQ Pup and HR Peg are both assigned to the disk 1 population.
        Their faint K and IRAS luminosities (table \ref{tab_Tc})
        are questionable but they are at the limit between O-rich and
        C-rich areas, like the other Tc S LPVs (fig. \ref{fig_Tc}).
        None of these stars show signs of dust emission
        (Jorissen and Knapp, 1998). They  are probably among the least
        massive stars in the disk population.

  \item $\chi$ Cyg, for which Jorissen and Knapp (1998) find questionable
        IRAS fluxes, presents no peculiarities here. Indeed its 25-12
        color index is close to zero but our estimated
        IRAS luminosities indicate the presence of an envelope. It is
        assigned to the disk 1 population and its location in
        the diagrams is compatible with a star of large initial mass in
        transit between O-rich and C-rich phases, after a recent
        dredge-up.

  \item R And can at first seem enigmatic. Assigned to the old disk 
        population, it is the most luminous star our sample in 12 and 25
        bands for its K luminosity and it has almost the smallest 25-12 index
        (see figure \ref{fig_Tc}). Our
        classification is probabilistic and thus a few stars can be
        misclassified, but this does not seem to be the case for R
        And. Indeed, the observed 25-12 color index is -0.9, in  
        agreement with that estimated from the IRAS absolute magnitudes
        (-0.8).

        This star probably has a very thick envelope and in the diagrams 
        it is close to the OH LPVs belonging to the old disk population (see
        figure \ref{fig_OH}). A mass close to the limit at which an O-rich
        LPVs can become either an OH emitter or a carbon star may account 
        for its characteristics.

\end{itemize}

\subsubsection{R Hor and Tc enrichment} \label{sec_RHor}

R Hor was found to be Tc enriched by Little et al. (1987) but not confirmed
as such by Van Eck (1999) and it is the only Tc star assigned to 
the extended disk
population. A Bayesian classification process can lead to some
misclassification, but R Hor is at the limit of the
O-rich LPVs area, close to two C-rich LPVs (RS Lup and V CrB)
(see table \ref{tab_pec}). Thus, its
Tc enrichment is questionable.

Another point is the way of enrichment in Tc and in C of such a deficient
and low mass star.  V CrB was reported as probably "metal poor" by Hron et
al. (1998) from ISO data.  This agrees with our assignation to 
ED (table \ref{tab_pec}), which seems a
priori doubtful for a Carbon star.

If this is confirmed, it would be an interesting constraint to
evolutive models.

\subsubsection{non-Tc S-type stars} \label{sec_noTcS}

Some of the non-Tc S stars in our sample can be extrinsic S stars. These
stars are not enriched in s-elements by internal nucleosynthesis and
dredge-ups but by mass transfer from a more evolved companion. This is
probably the case of X Aqr, BD Cam, V Cnc and SX Peg, for which a
duplicity flag is given in the HIPPARCOS catalog. Except for X Aqr, they
are in figure \ref{fig_Tc} at the lower limit in K of the old disk
population with thin envelope (ODf), to which they are assigned.
Moreover, the four of them have a K luminosity under -6.5 mag i.e. below
the threshold of thermal pulses on AGB, confirming the Van Eck's (1999)
result.

There are three other non-Tc S stars in our sample. They 
are the least luminous in 12 and 25 bands and have the
largest 25-12 index among stars with their K absolute magnitudes. AD Cyg
is assigned to the bright disk and it is more luminous than all C stars. It
is a massive star and we can assume that its evolution along the AGB is
very rapid. On the other hand, in the (K,25-12), (K,12), and (K,25) planes,
R Lyn and GZ Peg are close to the line on which AD Cyg and three extrinsic
S stars assigned to disk 2 population with a thin envelope (D2f) (BD Cam,
V Cnc, SX Peg) are located. These locations seem to confirm the non-Tc
character of these stars, which are probably extrinsic S stars. However,
given the
difficulties in detecting duplicity, we failed to confirm this result.\\

Although the number of extrinsic S stars in our sample prevented us 
from reaching
any definitive conclusion, these results suggest that an extrinsic
S enrichment can accelerate the evolution along the AGB 
 with formation of a circumstellar envelope closer to a carbon
than to a silicated composition, before any enrichment by the star's own
nucleosysthesis and dredge-ups.

\subsubsection{OH stars} \label{sec_OHstars}

Some O-rich LPVs are maser emitters on the radio
frequencies corresponding to OH bands, the masers being pumped by infrared
photons.

These stars emit at the principal frequencies of the main transition
(1665/67 MHz) and some also emit at 1612 MHz. The star is classified as OHII
or OHI according to a stronger or a lower emission at the secondary
frequency with respect to the main band.  All of them are O-rich LPVs.

A systematic research of OH masers for stars in the solar neighbourhood
has been carried out by Sivagnanam et al. (1989 and 1990), Lewis et al.
(1995)  and Szymczack et al. (1995).

\begin{figure*}

\centerline{\psfig{figure=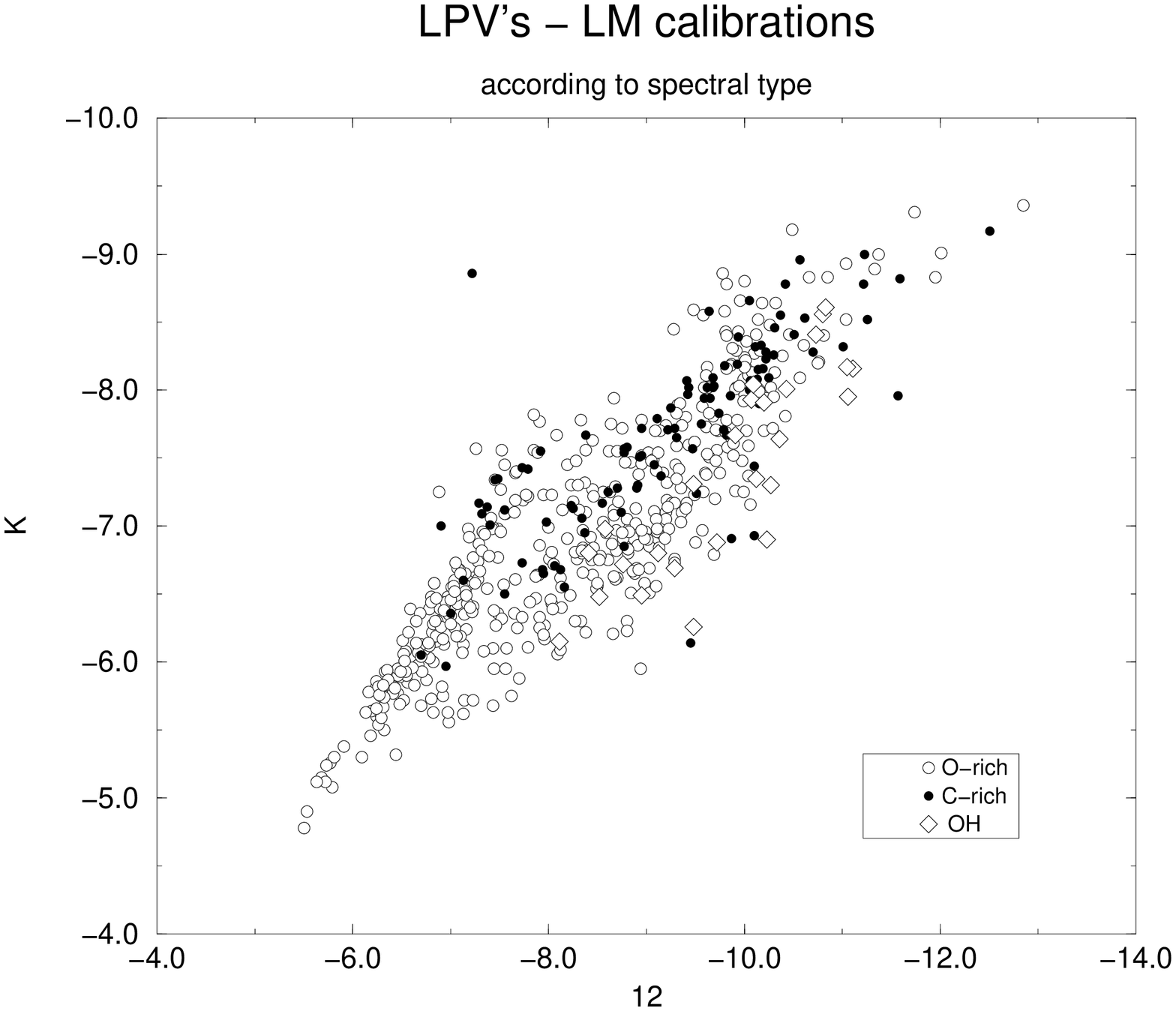,height=8cm ,angle=-90}
            \psfig{figure=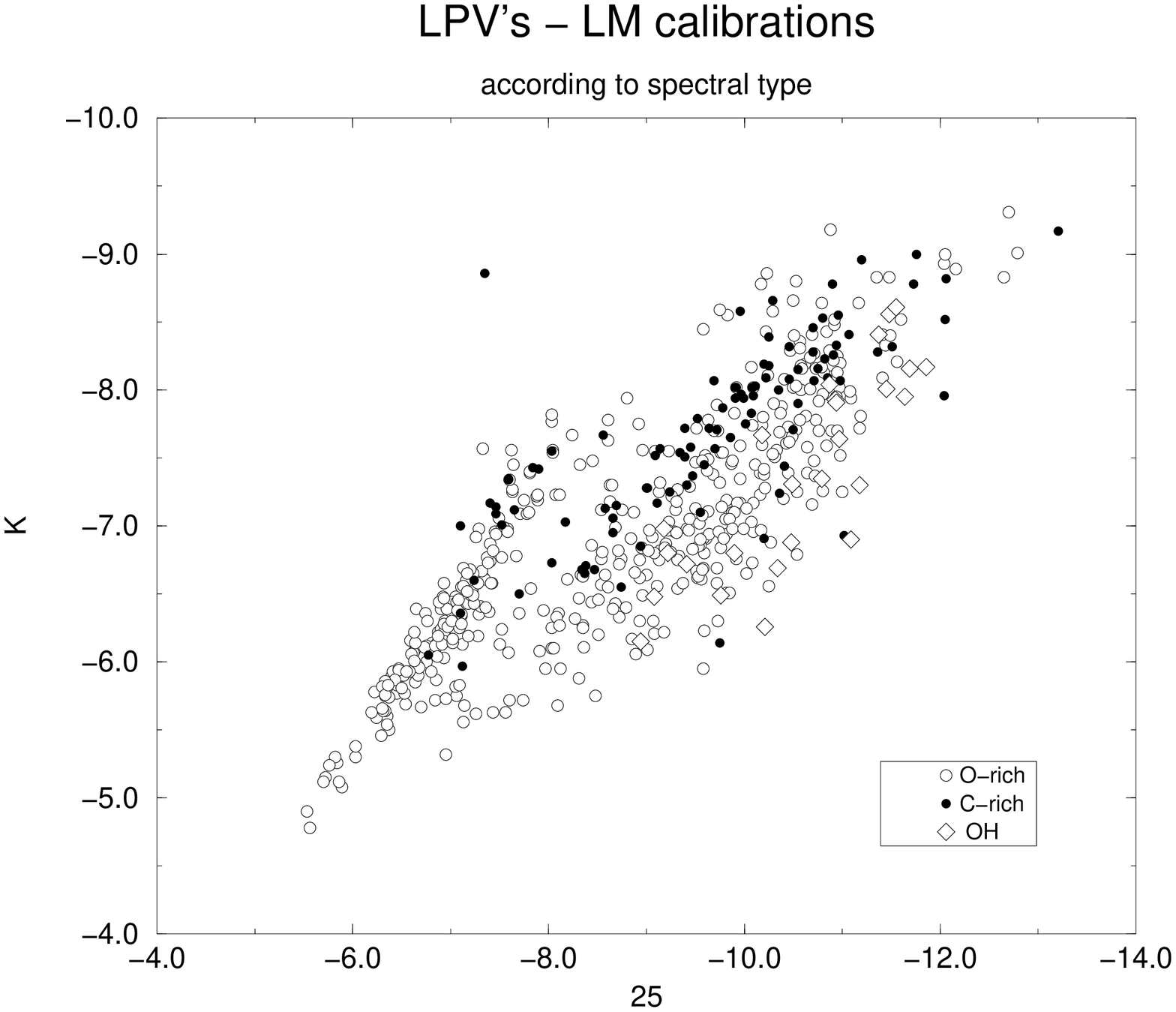,height=8cm ,angle=-90}}

\centerline{\psfig{figure=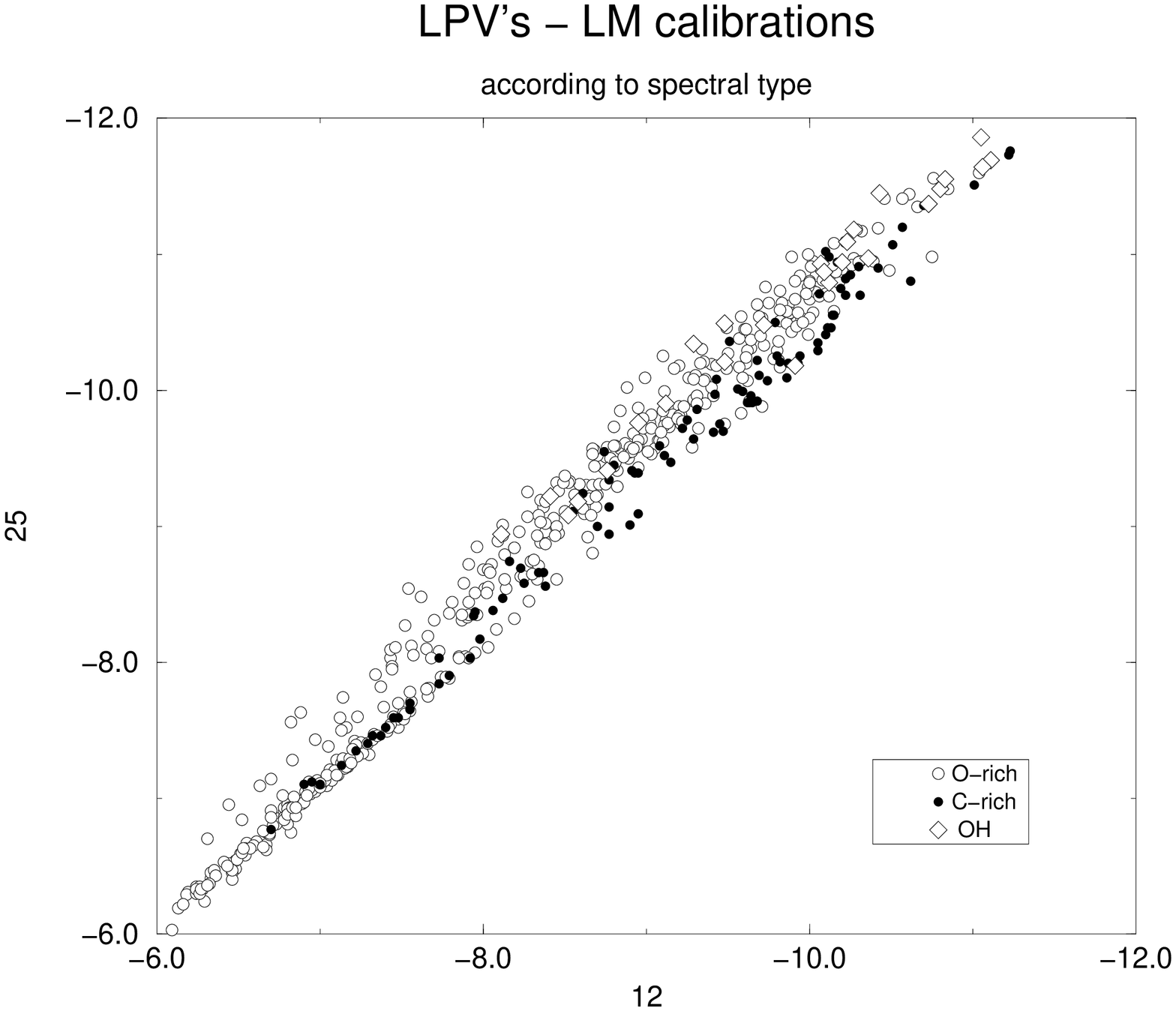,height=8cm ,angle=-90}
            \psfig{figure=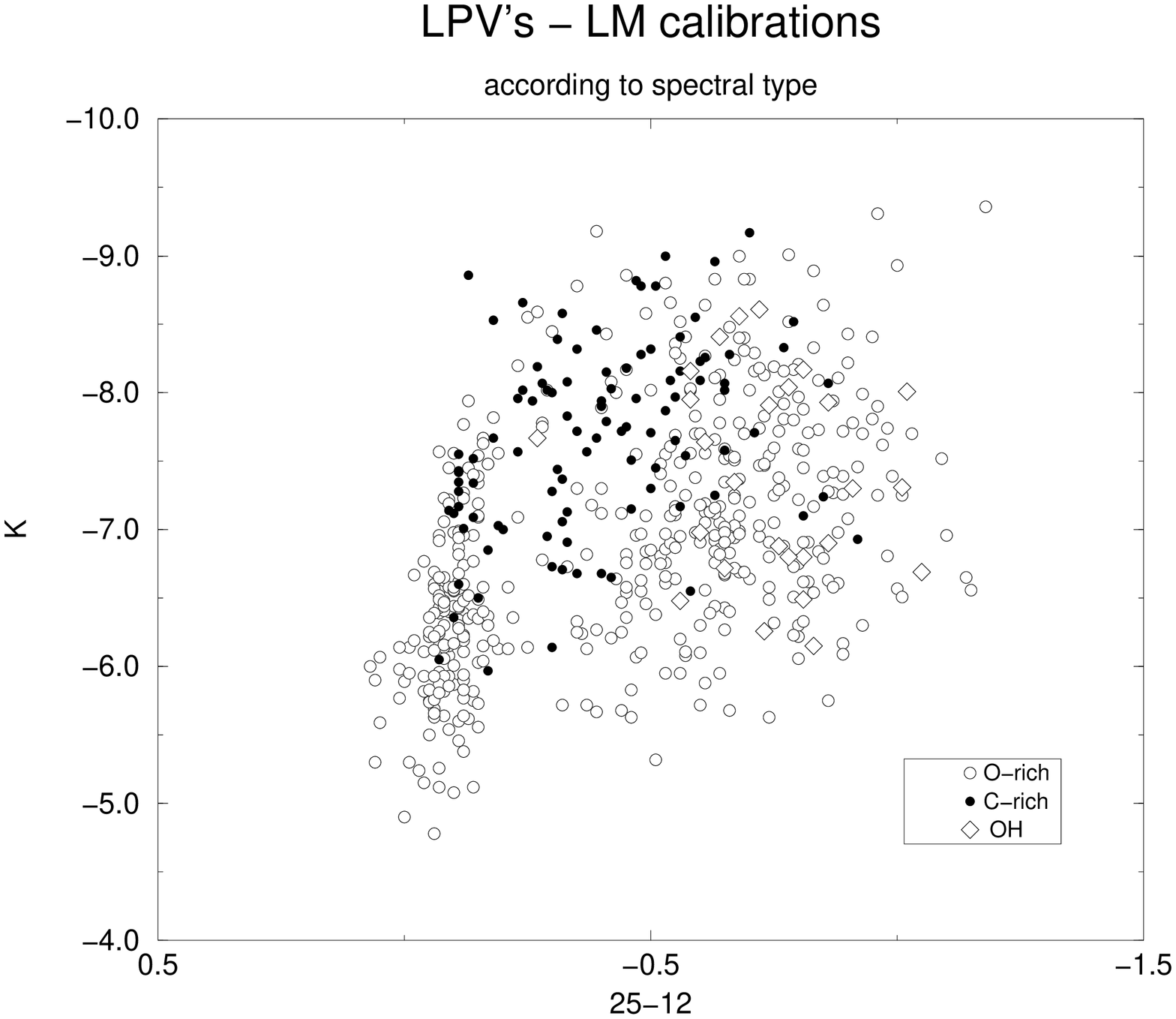,height=8cm ,angle=-90}}

\caption{Distribution of the individual estimated K, 12 and 25 luminosities and 
IRAS colors of OH stars compared to O and C-rich LPVs.
}

\label{fig_OH}

\end{figure*}

\begin{table*}

\caption{Individual K,12 and 25 luminosities, with assigned crossing K(IRAS) 
group of OH maser emitters Mira.
 }

\label{tab_OH}
\begin{center}
\begin{tabular}{|l|rr|r|r|rrr|r|}
   \hline
HIP id & \multicolumn{2}{c}{name} & types & group & K & 12 & 25 & pec \\
   \hline

 11350 & R     & Cet &    MO &  ODb &   -6.69 &   -9.29 &  -10.34 & OHI \\ 
 19567 & W     & Eri &    MO &   D2b &   -6.26 &   -9.48 &  -10.21 & OHI \\ 
 21766 & R     & Cae &    MO &     D1 &   -8.56 &  -10.80 &  -11.48 & OHI \\ 
 25673 & S     & Ori &    MO &   D2b &   -7.91 &  -10.20 &  -10.94 & OHI \\ 
 26675 & RU    & Aur &    MO &  ODb &   -6.90 &  -10.23 &  -11.09 & OHI \\ 
 27286 & S     & Col &    MO &   D2b &   -8.04 &  -10.09 &  -10.87 & OHI \\ 
 36669 & Z     & Pup &    MO &     D1 &   -8.17 &  -11.05 &  -11.86 & OHI \\ 
 40534 & R     & Cnc &    MO &   D2b &   -8.61 &  -10.83 &  -11.55 & OHI \\ 
 47066 & X     & Hya &    MO &  ODb &   -6.48 &   -8.52 &   -9.08 & OHI \\ 
 47886 & R     & LMi &    MO &     D1 &   -8.41 &  -10.73 &  -11.37 & OHI \\ 
 48036 & R     & Leo &    MO &     D1 &   -7.67 &   -9.91 &  -10.18 & OHI \\ 
 58854 & R     & Com &    MO &    D1 &   -6.80 &   -8.41 &   -9.22 & OHI \\
 67626 & RX    & Cen &    MO &  ODb &   -6.80 &   -9.12 &   -9.90 & OHI \\ 
 69346 & RU    & Hya &    MO &  ODb &   -6.49 &   -8.95 &   -9.76 & OHII \\ 
 69816 & U     & UMi &    MO &     D1 &   -6.98 &   -8.58 &   -9.18 & OHI \\ 
 70669 & RS    & Vir &    MO &    D1 &   -7.30 &  -10.27 &  -11.18 & OHII \\ 
 74350 & Y     & Lib &    MO &  ODb &   -6.15 &   -8.11 &   -8.94 & OHI \\ 
 75143 & S     & CrB &    MO &   D2b &   -8.01 &  -10.43 &  -11.45 & OHII \\ 
 75170 & S     & Ser &    MO &    D1 &   -6.72 &   -8.76 &   -9.41 & OHI \\ 
 79233 & RU    & Her &    MO &     D1 &   -7.93 &  -10.07 &  -10.93 & OHI \\ 
 80488 & U     & Her &    MO &   D2b &   -7.95 &  -11.06 &  -11.64 & OHII \\ 
 91389 & X     & Oph &    MO &     D1 &   -8.16 &  -11.11 &  -11.69 & OHI \\ 
 93820 & R     & Aql &    MO &  ODb &   -7.31 &   -9.48 &  -10.49 & OHII \\ 
 98077 & RR    & Sgr &    MO &   D2b &   -7.64 &  -10.36 &  -10.97 & OHI \\ 
 98220 & RR    & Aql &    MO &  ODb &   -6.88 &   -9.72 &  -10.48 & OHII \\ 
114114 & R     & Peg &    MO &   D2b &   -7.35 &  -10.12 &  -10.79 & OHI \\ 
\hline
\end{tabular}
\end{center}
\end{table*}

Individual estimated absolute magnitudes and assigned groups of OH stars
in our sample are given in table \ref{tab_OH}. Figure \ref{fig_OH} shows
their location in the distributions of the various absolute magnitudes.  
The distinction between OHI and OHII is not useful for our
purposes because no difference was found between them in our analysis. \\

As expected, no OH star was found with a thin envelope, i.e.
assigned to an f group. At a given K, they are the brightest in the 12
and 25 bands and all (except R Leo) have a 25-12 index corresponding to a
thick circumstellar envelope.  Their K luminosity distribution is the same
as that of non-OH O-rich LPVs. However, our previous V calibration
(Mennessier et al., 1999)  indicates the extent to which the presence of a thick
envelope induces an absorption in the visible range and confirms that
after a new growth of the envelope the star becomes fainter and fainter
in the visible range and turns into an OH-IR source. All these results agree
with the current model of OH sources: a maser emission pumped by photons of
an infrared thick envelope that depends on the mass of the star and the
mass-loss ratio.\\

  Finally, the kinematic assignation of OH emitters also provides
  information about these stars.
Ten stars were
assigned to the disk 1 population and they can thus be considered LPVs with
massive progenitors. However we found that this population group is rather
attractive for C-rich stars and repulsive for O-rich stars (Paper I). 
These ten OH
emitters have a K luminosity brighter than about -8 mag. They are located
in the same area as the Tc S LPVs in the (K,25-12), (K,12)  and (K,25)
planes. Thus such stars probably have a mass at the limit of the capability to
be sufficiently enriched in Carbon by successive dredge-ups.  As
discussed in Sec. (\ref{sec_HBB}) some of them could be Hot Bottom Burning
candidates.

The other OH sources of the sample, assigned to the disk 2 or old disk
population, are less massive.

\section{How to explain the gap for observed O-rich LPVs} \label{sec_gap}
 
As seen in Sect.\ref{sec_otoc} the distribution of O-rich LPVs is 
clearly bimodal in the diagrams of figure \ref{fig_Kcolor}, separating 
shell and no-shell stars. This gap may a priori be formed by several  
scenarios of the circumstellar shell formation, but one must not forget 
that the sample selection can also induce such an effect. Therefore,
we must take both physics and sampling into account when examining the  
validity of the proposed hypotheses.

\subsection{Several hypotheses} \label{sec_sevhyp}

Our sample is composed of variable stars observed by HIPPARCOS, i.e., 
selected from criteria based on visual magnitude.  
  As seen in Paper I, this selection effect is the most prominent in 
  comparison with the other ones: availability of K magnitude and IRAS 
  detection. 
A priori and with regard to the selection effects,
we propose three possible explanations for the striking gap observed
for O-rich LPVs in figures \ref{fig_Kcolor}, \ref{fig_K12} and
\ref{fig_K25}:

\begin{enumerate}

  \item During its evolution along the AGB, the star presents two stages
        of variability. Initially it is irregular or semi-regular of type b,
        then the pulsation stops and thereafter the star becomes variable
        for a second time as a semi-regular of type a or Mira and its 
        circumstellar envelope grows.

  \item The first thermal pulses induce an irregular variability of the star.
        Thereafter, the pulsation becomes more regular and a significant mass
        loss is the source of the circumstellar envelope formation. Owing to the
        composition of the silicated envelope, the 12 and 25 (more 25 than 12)
        luminosities suddenly and strongly increase. After this rapid phase the
        evolution quietly continues.

  \item At the beginning of the thermal pulses the SRb or irregular variables
        slowly develop a circumstellar envelope that grows and becomes 
        sufficiently thick to make the star undetectable in V
        magnitude. The envelope expands and becomes more
        transparent and thus the star becomes again visible in the visual
        magnitudes.

\end{enumerate}

Obviously, reality may be more complex than these three proposed
scenarios, and other explanations based on other results
may be suggested. 
  The following section is dedicated to the analysis of the three above proposed hypothesis.

\subsection{Consequences of each hypothesis} \label{sec_eachhyp}

First of all, {\it The first hypothesis is not realistic}. It has no
convincing physical justification and no such behaviour is found in the
models. We will thus discard it immediately.

\begin{figure*}

\centerline{\psfig{figure=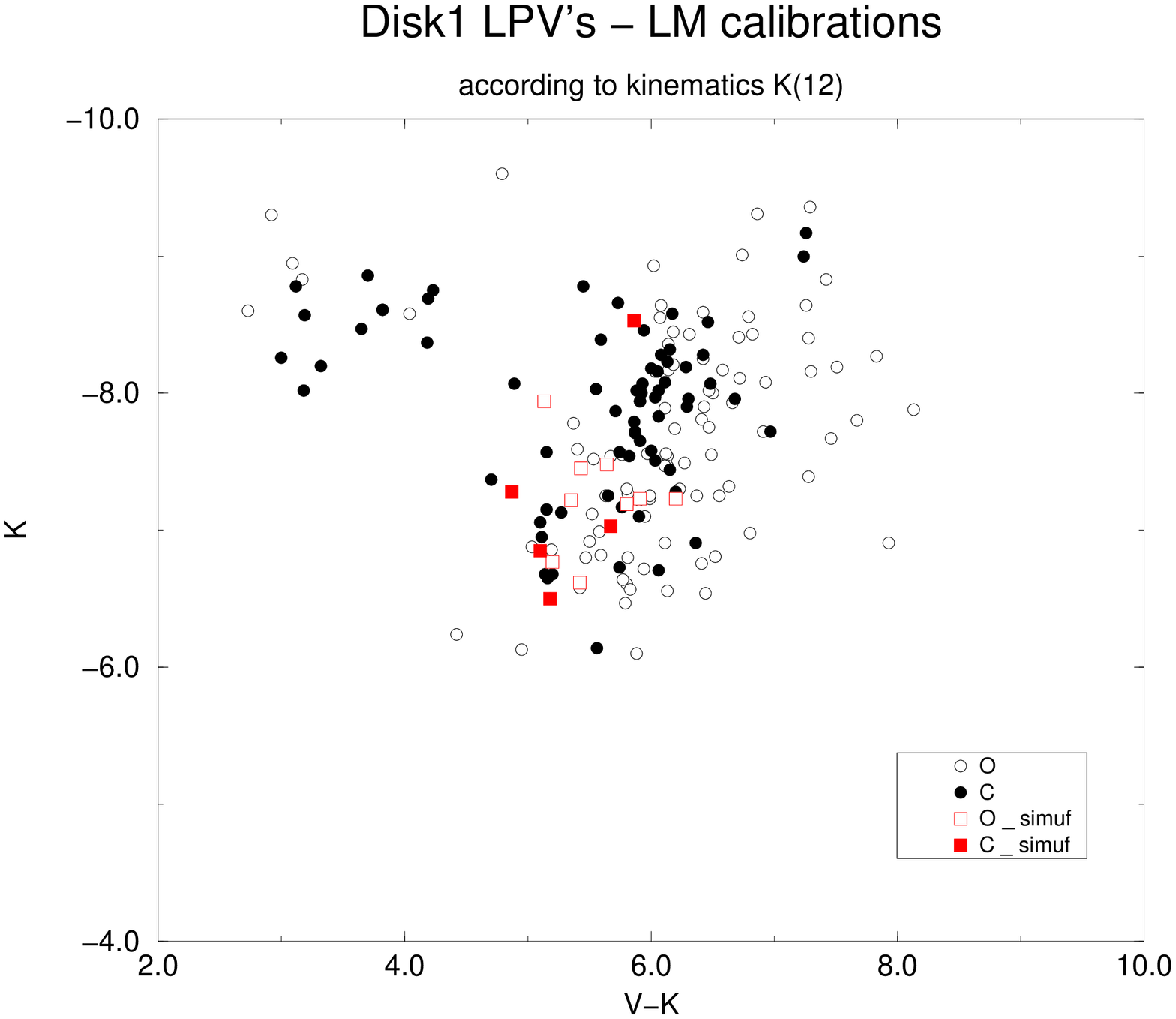 ,height=8cm ,angle=-90}
            \psfig{figure=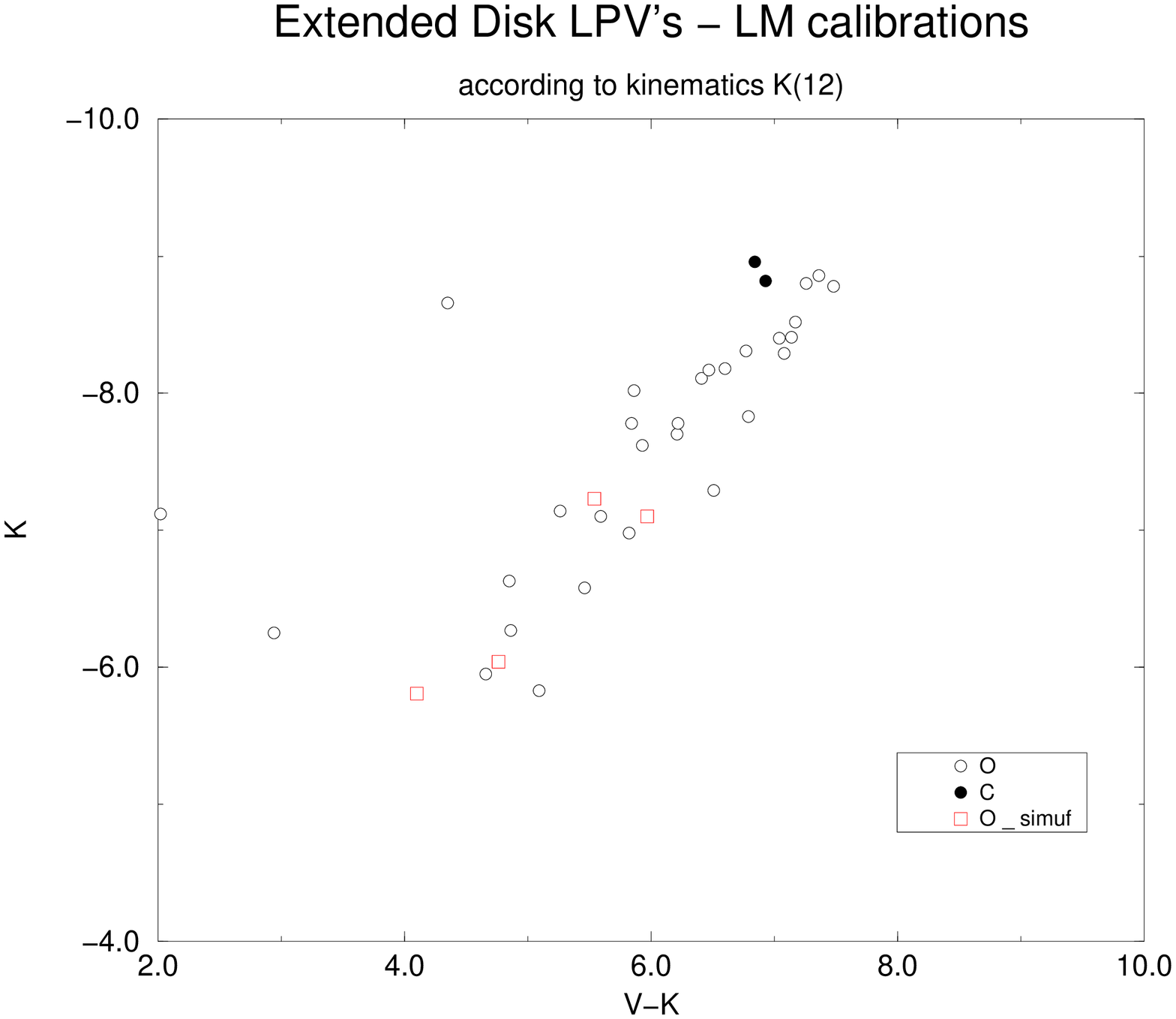,height=8cm ,angle=-90}}

\centerline{\psfig{figure=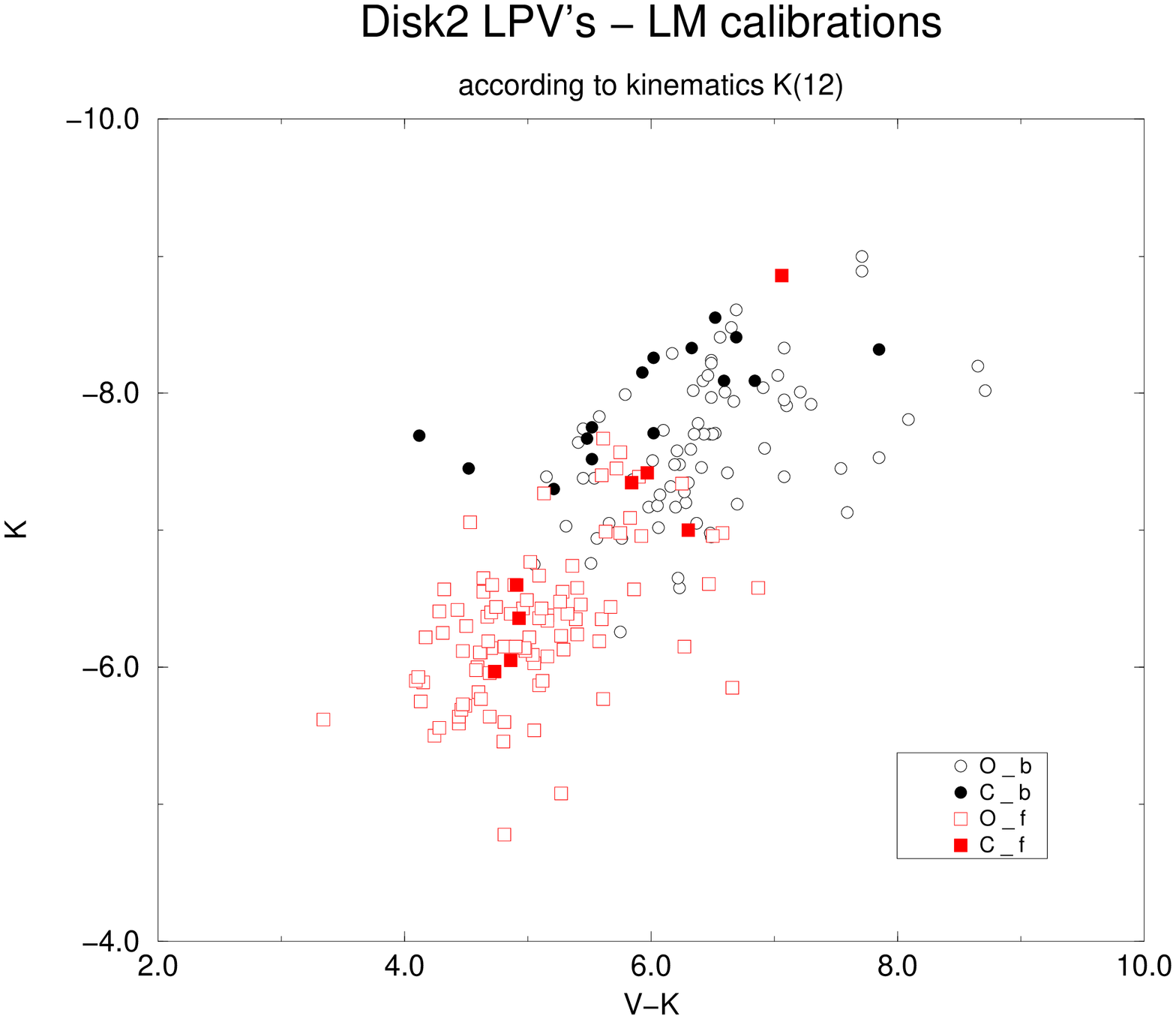 ,height=8cm ,angle=-90}
            \psfig{figure=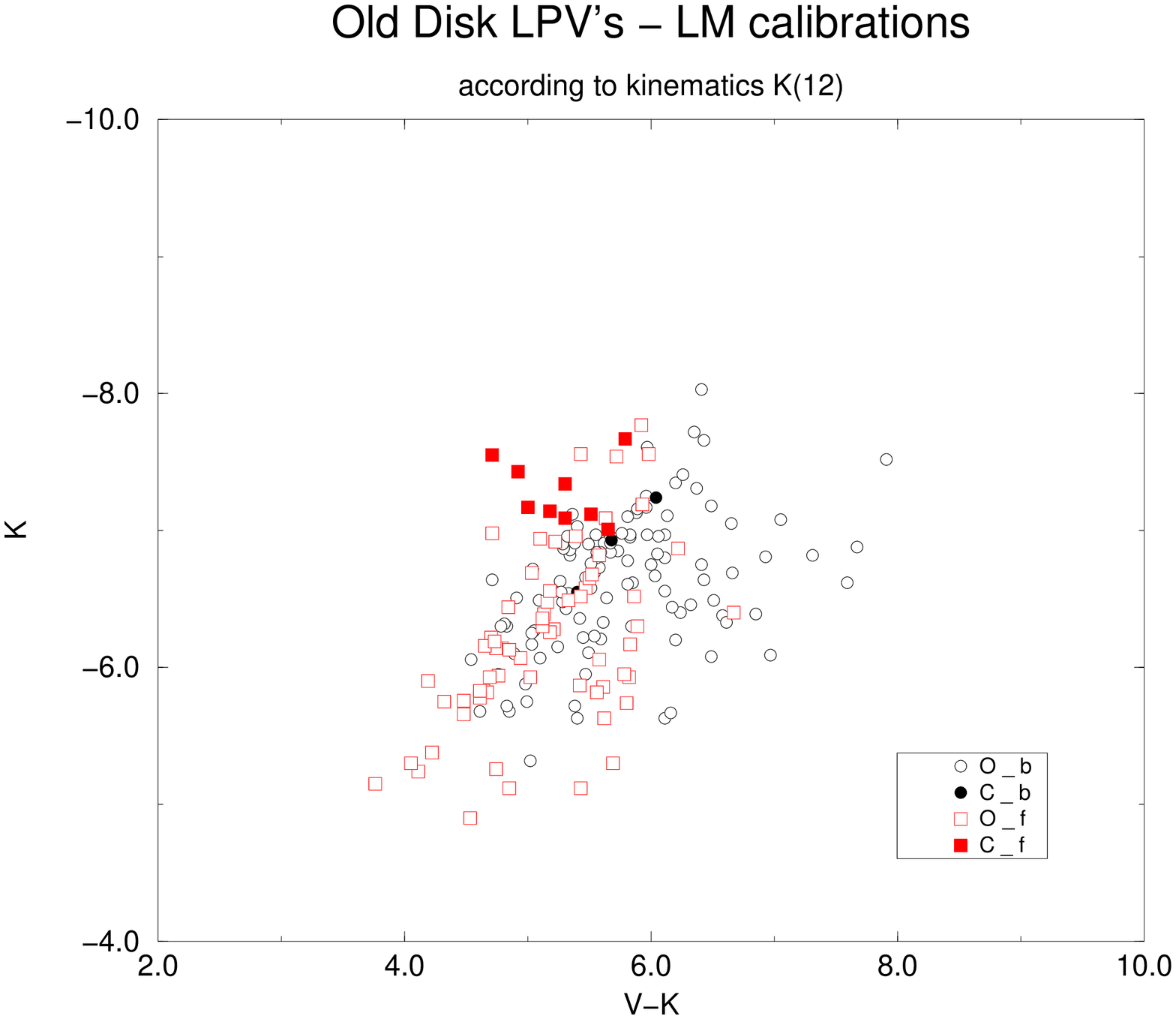,height=8cm ,angle=-90}}

\caption{Distribution of the individual estimated K luminosities and V-K
indices according to spectral types and indication of the envelope
thickness by separating stars assigned to f group or with a 25-12
index close to 0 (simuf).  Each figure corresponds to a kinematical group.
}

\label{fig_K-VK}

\end{figure*}

The other two are plausible. Let us examine their coherence using
Figure \ref{fig_K-VK}, which shows the distributions of the individual
estimated K absolute magnitudes and V-K indices from the V and K
luminosities estimated for each group.  Figure \ref{fig_K-VK} also
distinguishes the spectral types and the envelope thickness. A small
number of stars assigned to the disk 1 or extended disk population have a
25-12 index close to 0. They are marked in the figure \ref{fig_K-VK} as
"simuf" type points.

In the case of the third hypothesis, when the circumstellar envelope grows
more luminous in K, higher values of V-K are initially found; then V-K
decreases when the envelope expands. The star crosses back over the limit of
visibility, being more luminous in K and with a larger V-K index.

In the case of the second hypothesis, the change in the infrared fluxes is
rapid and there is no reason for a corresponding drastic change in K
luminosity, but the absorption in V band suddenly increases. Thus, the
stars belonging to the b group may be only visible (i.e. in our sample)  
from a more luminous lower limit in K than that of the f group.
Briefly, in the second hypothesis, from one side to the other of the gap,
the star has the same K luminosity but the K distribution of the sample
after the gap can be truncated for faint values because of an increase in
V-K. In the third hypothesis the gap corresponds to a time for which the 
star is invisible. Thereafter it is brighter in K.\\ 

  At this stage neither hypotheses 2 nor 3 can be rejected.
The data available are too scarce to decide between them. The reality may be 
a mixture of both, but as long as these two hypotheses are considered, 
it is obvious that the observed gap results from a circumstellar phenomenon. 
On the contrary, a discontinuity in the stellar evolution along the AGB can be 
excluded.

\subsection{Differences according to the galactic populations }
\label{sec_difevol}
 
 The (K,12) and (K,25) diagrams (fig. \ref{fig_K12} and \ref{fig_K25}) 
 allow us to extend our analysis. They show the extent to which the evolution
 depends on the galactic population, i.e. on the initial mass and
 metallicity. We remark that the faint K luminosity
 truncation of the sample of stars with a thick envelope 
 assigned to disk 2 population (D2b) is far
 lower than the one assigned to the old disk population (ODb).
 This difference between disk 2 and old disk populations 
 {\it favours the second hypothesis}.  Indeed such a difference is difficult to
 explain by an individual increase in K during the time of
 invisibility of the star assumed by the hypothesis 3. 
 On the other hand, a less massive star has a less efficient mass loss 
 with a formation of a less thick
 envelope and is thus less absorbed in the V band.  
 Therefore, in the case of  hypothesis 2,  at a given K for a thin envelope 
 star (f), 
 old disk population LPVs have a larger probability  
 than disk 2 population stars  to be in our V-selected sample just after 
 the circumstellar envelope formation. \\

The location of carbon stars also reinforces hypothesis 2. 
Indeed, a few C-rich LPVs have a 25-12 index close to zero but they 
cannot be stars 
just reaching the AGB. This is obvious when we examine 
the figure \ref{fig_Kcolor} except
for the 4 carbon stars belonging to the disk 2 population fainter than -7
mag. However, if we consider these 4 stars as 
having evolved from O-rich LPVs
belonging to the truncated faint part of the b distribution, everything is
consistent.

\begin{table*}

\caption{Individual K,12 and 25 luminosities, with assigned crossing
K(IRAS) group, spectral (O=O-rich, C=C-rich ,S=S spectral type star) and
variability (M=Mira,SR=semi regular,L=irregular) types, possible
specificity (Tc=Technetium star, OH=OH star, 
BD=bright galactic disk star, He=star in He-shell flash) and IRAS spectrum.}

\label{tab_pec}
\begin{center}
\begin{tabular}{|l|rr|r|r|rrr|r|r|}
   \hline
HIP id & \multicolumn{2}{c}{name} & types & group & K & 12 & 25 & pec & IRAS sp.\\
   \hline

 13502 & R     & Hor &    MO &   ED &   -8.40 &  -10.81 &  -11.49 & Tc? & \\
 70339 & RS    & Lup &    LC &   ED &   -8.96 &  -10.57 &  -11.20 &  & \\ 
 77501 & V     & CrB &    MC &   ED &   -8.82 &  -11.59 &  -12.06 &  & \\
       &       &     &       &        &      &        &      & & \\    
  1834 & T     & Cas &    MO &   D2b &   -9.00 &  -11.37 &  -12.05 & & 1 15\\
  7139 & IM    & Cas &  SROb &     D1 &   -8.95 &  -11.43 &  -12.36 & BD & 1 15\\ 
  7598 & V539  & Cas &    LO &     D1 &   -9.30 &  -11.50 &  -12.43 & BD & 1 16\\ 
 12302 & YZ    & Per &  SROb &     D1 &   -8.83 &  -11.70 &  -12.91 & BD & 1 27\\ 
 15530 & UZ    & Per &  SROb &     D1 &   -8.93 &  -11.04 &  -12.04 & & 1 25\\ 	
 46502 & Y     & Vel &    MO &     D1 &   -9.01 &  -12.01 &  -12.79 & & 1 23\\  
 69754 & R     & Cen &    MO &     D1 &   -9.60 &  -12.18 &  -12.79 & BD,He & 1 22\\ 
 87668 & V774  & Sgr &    LO &     D1 &   -9.36 &  -12.85 &  -14.03 & & 1 29\\
116705 & SV    & Cas &  SROa &     D1 &   -9.31 &  -11.74 &  -12.70 & & 1 25\\  
       &       &     &       &        &      &        &      & & \\  
 91389 & X     & Oph &    MO &     D1 &   -8.16 &  -11.11 &  -11.69 & OH & 1 15\\
       &       &     &       &        &      &        &      & & \\  
 32627 & V613  & Mon &   SROb &    D1 &  -8.60 & -10.07 & -10.31 & BD & \\
 89980 & V4028 & Sgr &   SROq &     D1 &   -8.58 &   -9.80 &  -10.29 & BD &  \\
       &       &     &       &        &      &        &      & & \\  
 65835 & R     & Hya &    MS &    D2b &   -8.52 &  -11.04 &  -11.60 & Tc,He & \\
 93820 & R     & Aql &    MO &    ODb &   -7.31 &   -9.48 &  -10.49 & OH,He & \\
  \hline
  \end{tabular}
  \end{center}
\end{table*}

\section{Peculiar evolution of bright massive LPVs} \label{sec_pecstars}
 
 The end of the evolution of massive LPVs on AGB induces complex  
 phenomena and deserves a more detailed study, presented in this section.

\subsection{Luminosity boundary of C stars}
 
 The luminosity boundary of carbon stars is an important constraint 
 for the models. We find the brightest C stars  around K=-9.2, 
 independent of metallicity (fig.\ref{fig_Kcolor}), in agreement with 
 the theoretical boundary: $M_{bol}$=-6.4 (Boothroyd et al.,1993). 

 However, in our individual estimates of absolute magnitude some
 O-rich LPVs are brighter in K than the brightest C stars (see table \ref{tab_pec}). 
 This could be due to Hot Bottom Burning (HBB) that prevents carbon 
 star formation. One of them, R Cen, is even brighter than the $3\sigma$ 
 upper limit of the AGB population  from our calibration
 (K=-9.4 for the disk 1 population, Paper I). 
 The properties of this exceptional star will be a guide  to investigate 
 the massive bright LPVs. 

\subsection{R Cen and Hot Bottom Burning} \label{sec_HBB}

R Cen has a very long period (more than 500
days) and its light curve presents a double maximum. It may have 
already changed from being a first overtone pulsator
to a fundamental one, but this assumption does not agree with
pulsation models (Ya'ari and Tuchmann, 1996) for a high luminosity star
with a period around 500 days.  Moreover, the hypothesis that R cen is a
first overtone pulsator accounts for the peculiar shape of its light
curve by a resonance phenomenon with the ratio P1/P3 close to 2 (Barthes,
1998).  This author suggests that this peculiarly massive star (more than
$3 {\cal M}_{\sun}$ and maybe $5 {\cal M}_{\sun}$)  is a candidate
star in the Hot Bottom Burning phase. Indeed this phenomenon stops the
Carbon-enrichment of the surface. This agrees with the fact that this
star is nearly the brightest in K,12 and 25 luminosities.\\

Two other candidates are proposed by Barthes (1998): T Cas and X Oph.  In
Paper I, we assign both these stars to the disk  population (see table
\ref{tab_pec}), in agreement with young and initially massive LPVs. Our
estimated luminosities confirm T Cas as an HBB
candidate. X Oph is more intriguing because its mass is probably 
at the lowest limit of carbon star formation (sect.\ref{sec_OHstars}).\\

Another peculiarity of R Cen is the appearance of the silicate band of its
ISO-SWS spectrum (Justtanont et al., 1998). Let us examine the O-rich LPVs
assigned to the disk 1 population with an estimated high luminosity. V774
Sgr and SV Cas have IRAS-LRS spectra (IRAS Science Team, 1986)  with a
similar appearance as that of R Cen; they also have a relatively early
spectral type (M5 and M6.5 respectively), and so they may also be HBB
candidates. IM Cas and V539 Cas have a high luminosity, an early spectral
type (M2)  and their IRAS-LRS spectra are close to that of R Cen one in the
same way as T Cas and X Oph. Another possible candidate could be Y Vel,
which has a late spectral type LPV (M8-M9.5).

At least YZ Per has an IRAS-LRS spectrum close to that of R Cen, an early
spectral type (M1-M3), but it is classified as a supergiant of class Iab
and so it may be at the most advanced end of the AGB or may be a post-AGB.\\
 
 IRAS luminosities and 25-12 indices of V4028 Sgr and V613 Mon indicate  
 that they probably are in the first stages on the AGB. 
 This is confirmed by the IRAS-LRS 
 spectrum of V4028 Sgr in which no SiO feature is present. Unfortunately 
 no IRAS-LRS spectrum of V613 Mon is available.

\subsection{R Cen and He-shell flash} \label{sec_He}

 Another interesting property of R Cen was studied by Hawkins et al. (2001) 
 in a recent paper. They present evidence of a steadily 
 decreasing period of R Cen from 550 to 505 days during the last 50 years.
 They suggest that it is caused by a He-shell flash, in a similar way to 
 R Hya, R Aql, W Dra and T UMi (Wood and Zarro, 1981).
 
 W Dra and T UMi have no available HIPPARCOS data and so we could not 
 estimate their luminosities. R Aql is an OH emitter and  a  
 remark similar to X Oph (sect.\ref{sec_HBB}) applies. 
 
 Wood and Zarro (1981) deduce a value between -5.3 and -5.5 mag. 
 for the bolometric luminosity of R Hya, in agreement
 with our estimation (K=-8.58).
 They also find that the time scale for the period change 
 of R Hya corresponds to a longer time after the 
 maximum luminosity of the He-shell flash than that of R Aql. This could 
 explain the Tc enrichment of R Hya and its  M6-M9S spectral type.\\
 
 The period change for R Cen is steeper than for R Hya and 
 thus Hawkins et al. (2001) give two possible explanations: either R Cen is 
 in a stage right after the beginning of the flash, with a 
 total mass less than 2-3 ${\cal M}_{\sun}$ or it is in a stage right after 
 where the luminosity of the flash reaches the stellar surface 
 with a much larger 
 range of allowed stellar mass. 
 We find that R Cen is the most K luminous LPV  but 
 its 25-12 index is this of an S star, at the limit between O and C-rich 
 LPVs, and is assigned to BD, the group of the most massive stars.
 Therefore, our results strongly 
 favor the second possibility: R Cen is in a stage right after 
 the luminosity of the flash reaches the stellar surface.
 
 Furthermore the He-shell flash enhances the efficiency of the third dredge-up 
 (Herwig, 2000), that can explain the very peculiar location, compared 
 to the one of 
 that HBB candidates, of R Cen 
 in the diagram (K,25-12). This so luminous O-rich LPV might  become 
 a carbon star exceptionally brighter than the usual  luminosity 
 limit accepted for these stars. A few such luminous carbon stars 
 exist in the Magellanic Clouds 
 as observed by van Loon et al. (1999b) and modeled by Frost et al. 
 (1998). 

\section{Conclusion} \label{sec_conclu}

Our results confirm that the AGB evolution depends on the initial mass of
the progenitor on the main sequence. The study of LPVs with peculiar
properties, often associated with transition states in the stellar evolution,
elucidates some points of the very complex evolution along the AGB.  
The simultaneous study of the behaviour of the circumstellar envelope provides
further information on the evolutive state of the stars along the AGB.
However, this study is mainly statistical and so
results for individual stars can be erroneous because the confidence
level of a probabilistic discrimination can never reach 100\%.\\

\begin{figure*}

\centerline{\psfig{figure=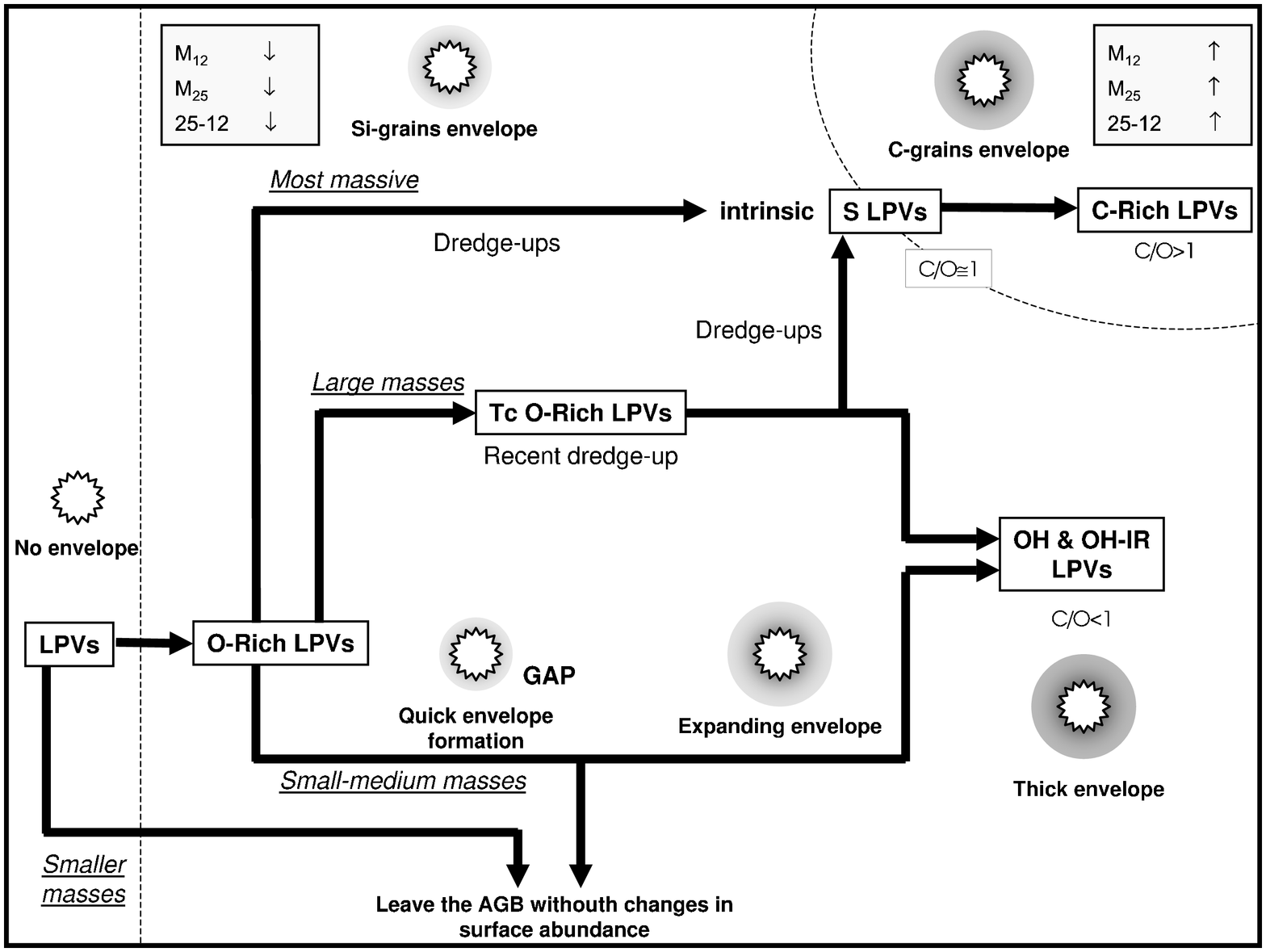,height=11cm ,angle=-90}}
 
\caption{Schematic illustration of the proposed stellar and circumstellar 
evolutive scenario}

\label{fig_schema}

\end{figure*}

The proposed evolutive scenario, schematically represented in figure
\ref{fig_schema}, is:

\begin{itemize}

  \item At the beginning of the variability phase, stars
        belonging to the disk 2 or old disk population (i.e., with 
        a not too large initial mass ${\cal M}_{ms}$) are  irregular L or
        semi-regular SRb O-rich variables. They begin to slowly  produce
        an envelope. Thereafter the envelope expands, and the IRAS luminosities         of the star rapidly grow brighter. They become O-rich
        SRa or Mira. Depending on  smaller or greater ${\cal M}_{ms}$,
        they can evolve either to OH emitters and finally OH-IR sources
        or, for the more massive, to be enriched in s elements and then in
        carbon by successive dredge-ups and finally become a C-rich
        Mira. The least massive stars leave the AGB with no special 
        transformation of their surface abundance.

  \item The disk 1 and bright disk population LPVs seem to very rapidly build a
        bright expanding envelope. They can also evolve to OH or C-rich stars. 
        Although this population is not attractive for O-rich LPVs, a tenth
        of OH Mira belong to it. They are probably not massive enough to be
        sufficiently Carbon enriched. A few may be HBB stars.

        If the mass of the star is high enough, after a number of dredge-ups 
        the external shells of the star can be enriched in Carbon. The C/O
        ratio becomes larger than 1. When C/O
        is around 1, the star is an S star. At the same time,  strong changes
        take place in the circumstellar envelope, which becomes dominated by
        carbonated grains, and the 25-12 index increases in conformity with
        the loop drawn in the IRAS color-color diagram, as predicted by Willems
        and de Jong (1988) and calculated by Chan and Kwok (1988).
        Our luminosity calibrations in 12 and 25 clearly show that
        the star becomes fainter in both luminosities but more in 25 than 
        in 12. The C-rich irregular and SRb stars seem to be the most evolved
        and massive
.
\end{itemize}

The study of LPVs enriched in Tc confirms that they are at different
stages along the AGB but after a recent dredge-up. It allows us to confirm
the location of the first dredge-up at $M_{bol}=-3.5$ and the quite early
operative third dredge-up on the TP-AGB\\

The no-Tc S-type LPVs (except R And), are faint in K, 12 and 25, and they
are confirmed as extrinsic S stars enriched not by their own
nucleosynthesis but by mass exchange from a more evolved companion. The
extrinsic enrichment in s-elements may accelerate the evolution along
the AGB and lead to the formation of an envelope closer to being
carbonated than silicated before any intrinsic enrichment by successive
dredge-ups.\\

 The examination of the brightest LPVs allows us to 
propose a list of stars with peculiar spectral, envelope and
luminosity properties that may be Hot Bottom Burning candidates. 
The most luminous of them, R Cen, a star in a He-shell flash, 
could become, before leaving AGB, a C-rich LPV brighter than the usual 
luminosity limit of carbon stars.

\begin{acknowledgements}

This study was supported by the PICASSO program PICS 348 and by the
CICYT under contract ESP97-1803 and AYA2000-0937. We thank 
N.Mowlavi and R.Alvarez for constructive remarks and 
A.Gomez and S.Van Eck for fruitful discussions of our preliminary results.

\end{acknowledgements}

\end{document}